\documentclass[useAMS,usenatbib]{mn2e}
\usepackage{aas_macros}
\usepackage[a4paper,centering, totalwidth=520pt, totalheight=700pt]{geometry}
\usepackage{graphicx}
\usepackage{amssymb}
\usepackage{amsmath}

\usepackage{color}

\newcommand{\Msun}{\ensuremath{M_{\odot}}}
\newcommand{\Mh}{\ensuremath{h^{-1}M_{\odot}}}
\newcommand{\Mhsq}{\ensuremath{h^{-2}M_{\odot}}}
\newcommand{\Mpch}{\ensuremath{h^{-1}{\rm Mpc}}}

\newcommand{\avg}[1]{\ensuremath{\left\langle \,#1\, \right\rangle}}

\newcommand{\der}{\ensuremath{{\rm d}}}

\newcommand{\eqn}[1]{equation~\eqref{#1}}
\newcommand{\eqns}[1]{equations~\eqref{#1}}

\newcommand{\ph}[1]{\phantom{#1}}

\newcommand{\be}{\begin{equation}}
\newcommand{\ee}{\end{equation}}
\newcommand{\Cal}[1]{\ensuremath{\mathcal{#1}}}

\definecolor{darkgreen}{rgb}{0,0.5,0}

\title[Conformity from correlating colour and concentration]
{Correlating galaxy colour and halo concentration: A tunable Halo Model of galactic conformity}
\author[A. Paranjape, et al.]{Aseem Paranjape\thanks{Email: aseem@iucaa.ernet.in}$^{1,2}$, Katarina Kova\v c\thanks{Email: kovac@phys.ethz.ch}$^{2}$, William G. Hartley\thanks{Email: hartleyw@phys.ethz.ch}$^{2}$ and Isha Pahwa\thanks{Email: ipahwa@aip.de}$^{1,3}$ \\
$^{1}$Inter-University Centre for Astronomy and Astrophysics, Ganeshkhind, Post Bag 4, Pune 411007, India\\
$^{2}$Institute for Astronomy, Department of Physics, ETH Z\"urich, Wolfgang-Pauli-Strasse 27, CH 8093, Switzerland\\
$^{3}$Leibniz-Institut f\"ur Astrophysik Potsdam (AIP), An der Sternwarte 16, D-14482 Potsdam, Germany
}
  
\begin{document}
\date{\today}
\pagerange{\pageref{firstpage}--\pageref{lastpage}} \pubyear{2015}
\maketitle
\label{firstpage}

\begin{abstract}
We extend the Halo Occupation Distribution (HOD) framework to generate mock galaxy catalogs exhibiting varying levels of ``galactic conformity'', which has emerged as a potentially powerful probe of environmental effects in galaxy evolution. Our model correlates galaxy colours in a group with the concentration of the common parent dark halo through a ``group quenching efficiency'' $\rho$ which makes older, more concentrated halos \emph{at fixed mass} preferentially host redder galaxies. We find that, for a specific value of $\rho$, this 1-halo conformity matches corresponding measurements in a group catalog based on the Sloan Digital Sky Survey. 
Our mocks also display conformity at large separations from isolated objects, potentially an imprint of halo assembly bias. A detailed study -- using mocks with assembly bias erased while keeping 1-halo conformity intact -- reveals a rather nuanced situation, however. At separations $\lesssim 4$Mpc, conformity is mainly a 1-halo effect dominated by the largest halos and is \emph{not} a robust indicator of assembly bias. Only at very large separations ($\gtrsim 8$Mpc) does genuine 2-halo conformity, driven by the assembly bias of small halos, manifest distinctly. We explain all these trends in standard Halo Model terms. Our model opens the door to parametrized HOD analyses that self-consistently account for galactic conformity at all scales.
\end{abstract}



\section{Introduction}
\label{sec:intro}
\noindent
The processes that cause star formation activity in galaxies to diminish or ``quench'' are of great interest in understanding galaxy evolution. A number of observational and (semi-)numerical studies \citep{vdb+08,rb09,ww10,vdw+10,hopkins+10,more+11,prescott+11,wtc12,wtcv13,cen14,hirschmann+14} have attempted to distinguish between the relative importance of galaxy mergers \citep{tt72} and wider halo-level physical processes such as ram-pressure stripping of gas \citep{gg72}, ``strangulation'' \citep{ltc80,bm00}, ``harassment'' \citep{moore+96}, etc. that may lead to quenching in group environments \citep[for a recent review, see][]{sd15}. Even where halo-level effects are indicated, there is ongoing debate over whether the location of a galaxy within its group (i.e. whether it is a ``central'' galaxy or a ``satellite'') is paramount \citep{baugh+06,peng+12,kovac+14}, or whether the nature of the group as a whole is more important in quenching its member galaxies \citep{osmond+04,vdb+08,carollo+14,hartley+15,klwk14}.

In this context, a particularly interesting phenomenon that has received attention recently is that of ``galactic conformity'' \citep{weinmann+06a}. This is the observation that satellite galaxies in groups whose central galaxy is quenched (or  red) are preferentially quenched, even when the groups are restricted to reside in dark matter halos of the same mass \citep[see][for an in-depth study extending this statement to a variety of measures of environment]{klwk14}. \cite{weinmann+06a} based these conclusions on a group catalog \citep{yang+05} constructed using galaxies in the Sloan Digital Sky Survey (SDSS). 
More recently, a similar effect has been observed by \citet{kauffmann+13} when studying the star formation rates (SFR) of SDSS galaxies at large spatial separations ($\lesssim4$Mpc) from isolated objects.

The observation of galactic conformity contradicts an assumption typically made in halo occupation models of galaxy distributions, namely, that galaxy properties such as luminosity and colour distributions are determined solely by the mass of the parent dark halo of the group to which the galaxy belongs. Modelling conformity then becomes important in order to better understand the galaxy-dark matter connection \citep{zhv14}, with consequences for both semi-analytical models of galaxy evolution \citep{wwdy13,kauffmann+13} as well as statistical studies aimed at constraining cosmological parameters \citep{vdb+13,more+15,coupon+15}.

It is tempting to ascribe a physical origin to galactic conformity, as \cite{hartley+15} do. Since the seminal work of \cite{fabian+03}, evidence for regulation of the hot gas temperature in clusters of galaxies, by the super-massive black hole (SMBH) at the centre of the brightest cluster galaxy, has been growing \citep{croton+06,zhuralavleva+14}. Although it is not clear whether a similar regulation exists for group-mass halos, a coupling of the star-formation properties of different group members via the hot gas content of their halo provides a natural explanation for conformity; a coupling that could even be established many Gyrs previously \citep{rawlings+04}. 
A coupling such as this may also exist due to re-processing of the gas content of individual galaxies in a group \citep{oppenheimer+10,birrer+14}.
However, this is far from a unique solution. We might equally speculate that the virial shock heating of infalling gas depend on the halo's location within the cosmic web, perhaps being more effective in filaments than in voids at low halo masses for instance. In each of these hypothetical cases the conformity signal found by \cite{weinmann+06a} and others is driven by a hidden variable (past/present SMBH activity or large scale environment), i.e. a quantity that was not accounted for in those studies. Clearly identifying the correct hidden variable(s), those that ``explain'' conformity, will lead us to a greater understanding of the physical nature of quenching in galaxy groups.

An obvious suspect as to the hidden variable, and therefore the possible origin of galactic conformity, is the assembly history of the parent halo. The phenomenon of halo assembly bias, in which low mass older halos tend to cluster more strongly at scales $\gtrsim10$Mpc than younger, more recently assembled halos of the same mass (while this trend reverses for more massive halos) is well established \citep{st04,gsw05,wechsler+06,jsm07,desjacques08,dwbs08,hahn+09,fm10}. Halo age also correlates well with other halo properties -- we will focus on halo concentration which correlates positively with age \citep{nfw97,wechsler+02} -- and halo assembly bias has been observed to extend to these properties as well; e.g., at fixed low mass, more concentrated halos typically cluster more strongly than less concentrated ones. 

Viewed from the present day, a halo that has a low late-time mass accretion (with respect to other similar mass haloes) would have a higher formation redshift and, statistically, a greater concentration. In relatively low-mass haloes the build up of dark matter mass and the accretion of baryons by the central galaxy are quite tightly coupled, because the time scale for cooling to offset gravitational heating is shorter than the free-fall time \citet{wr78}. The galaxies hosted by early-forming haloes would therefore experience a reduced supply of star-forming gas with respect to the larger galaxy population, and in the most extreme cases could result in a higher probability of being quenched. A plausible hypothesis, then, is that halo assembly bias affects galaxy formation and leads to a \emph{galaxy} assembly bias \citep[][]{tinker+12}, which leaves an imprint in the form of galactic conformity \citep{hwv15}. More specifically, according to this hypothesis galactic conformity both within individual groups \emph{as well as} at large scales may be explained if, at fixed halo mass and galaxy luminosity/stellar mass, quenched  or red galaxies preferentially reside in older (sub)halos. 

One point of interest in this regard is that the conformity signal observed by \citet{kauffmann+13} at scales $\lesssim4$Mpc is quite strong (roughly an order of magnitude difference in the median specific SFR of galaxies surrounding isolated objects that belong to the upper and lower 25th percentiles of specific SFR). Halo assembly bias, on the other hand, has been argued to be relatively unimportant in modelling, e.g., the dependence of galaxy 2-point correlation functions on properties such as luminosity, colour and SFR at similar scales \citep[e.g.,][]{ss09,deason+13}. It is therefore important to 
ask how assembly bias can be simultaneously consistent with these observations, or whether some other mechanism could generate the conformity signal.

In principle, such a test can be performed by embedding an assumed conformity-inducing correlation between galaxy and halo properties in an analytical Halo Model framework \citep[for a review, see][]{cs02}, since it is feasible to accurately model property-dependent clustering in this language \citep{wechsler+06}. However, comparing analytical calculations directly with observations can become quite involved in the presence of possibly complex selection criteria used in defining the observed sample and the signal itself. A more efficient method, then, is to use $N$-body simulations of dark matter and construct mock galaxy catalogs in which one has full control over switching the conformity signal on and off; these mocks can then be analysed in the same way as the observed sample for a fair comparison. Such mock catalogs have recently been presented by \cite{hw13,mly13,hearin+14a} who model galactic conformity by introducing a rank ordering of galaxy colours or SFR by suitably defined measures of (sub)halo age. The ``age matching'' mocks of \citet{hw13} and \citet{hearin+14a} have been shown to successfully reproduce a number of observed trends such as the 2-point correlation function and galaxy-galaxy lensing signal of SDSS galaxies \citep{hearin+14b}, the radial profile of the satellite quenched fraction in groups \citep{watson+15}, etc., in addition to containing a conformity signal.
The rank ordering in these mocks leads to a fixed strength of galactic conformity, however \citep[see, e.g., Figure 3 in][]{hwv15}. In the absence of conclusive evidence as to the nature of the hidden variable that explains conformity, it is then interesting to explore alternative algorithms, particularly those that implement a \emph{tunable} effect which can then be compared with observations. 

Another relevant issue is that it is important to be able to distinguish cleanly between conformity within individual groups and conformity effects due to spatial correlations between distinct halos \citep[these were respectively dubbed ``1-halo'' and ``2-halo conformity'' by][]{hwv15}. This is because the distinction between what is inside and outside a dark halo becomes fuzzy in any analysis that averages over a large range of halo masses. E.g., we would like to ask the following: out to what spatial separations might one see the effects of conformity in a model that has \emph{only} 1-halo conformity built in and \emph{does not} know about the large scale environmental correlations due to halo assembly bias? This might plausibly be the case in a halo-specific gas regulation mechanism that couples to star formation activity as mentioned above. This question is particularly interesting in the context of the apparent conflict in the expected strength of galaxy assembly bias and the observed strength of galactic conformity mentioned above. 

In this paper we introduce an algorithm that allows us the level of control needed to address the above issues. Our algorithm is a modification and extension of the one described in \citet[][hereafter, S09]{ss09} to model galaxy positions, luminosities and colours; we model galactic conformity by introducing a positive correlation between galaxy colour and the concentration of the parent dark halo \emph{of the group} to which the galaxy belongs (i.e., we work at the level of groups and not subhalos). The strength of this correlation is determined by adjusting the value of a parameter $\rho$ which we interpret as a ``group quenching efficiency'', the terminology being motivated by similar quantities studied by earlier authors \citep[see, e.g.,][]{vdb+08,peng+10,klwk14}. In order to answer the question regarding large scale effects raised above, we also explore a model where halo concentrations are \emph{randomized} among halos of fixed mass before correlating them with the galaxy colours, thereby erasing any large scale clustering of the colours due to halo assembly bias while keeping average group-specific properties (including 1-halo conformity) intact.

The paper is organised as follows.
In section~\ref{sec:algorithm} we describe the S09 algorithm, followed by our modifications for correlating galaxy colours with halo concentrations and calculating stellar masses.
In section~\ref{sec:simulations} we describe the $N$-body simulations on which our mock galaxy catalogs are based. Section~\ref{sec:data} describes the SDSS-based group catalog that we use for comparison with our mocks, together with various fitting functions derived from this sample which inform our mocks. 
Our results are described in section~\ref{sec:results}, followed by a discussion in section~\ref{sec:discuss}. 
We conclude in section~\ref{sec:conclude}.
We will use SDSS galaxy properties K-corrected to redshift $z=0.1$, usually denoted by a superscript, e.g., ${}^{0.1}r$ for the $r$ band. Throughout, we will denote $M_{{}^{0.1}r}-5\log_{10}(h)$ as $M_r$, ${}^{0.1}(g-r)$ as $g-r$ and quote stellar masses $m_\star$ in units of \Mhsq\ and halo masses $m$ in units of \Mh, where $H_0=100\,h\,{\rm km\,s}^{-1}{\rm Mpc}^{-1}$ is the Hubble constant. We use a flat $\Lambda$CDM cosmology with parameters $\Omega_{\rm m}=0.25$, $\Omega_{\rm b}=0.045$, $h=0.7$, $\sigma_8=0.8$ and $n_{\rm s}=0.96$, which are consistent with the 5-year results of the WMAP experiment \citep{hinshaw+09}. 


\section{Algorithm}
\label{sec:algorithm}
\noindent
Our basic algorithm is borrowed from S09 and is itself an extension of the algorithm described by \citet{sscs06} to include galaxy colours. The algorithm uses a Halo Occupation Distribution function \citep[HOD;][]{bw02} calibrated on SDSS luminosity-dependent projected clustering measurements to create a luminosity-complete mock galaxy catalog. We describe this algorithm and our modifications below.


\subsection{The S09 algorithm}
\label{sec:algo:subsec:S09}
\subsubsection{Luminosities, positions and velocities of centrals and satellites}
\noindent
The S09 algorithm explicitly implements the so-called central-satellite split. A fraction $f_{\rm cen}(<M_{r,{\rm max}}|m)$ of $m$-halos (i.e., halos with masses in the range $(m,m+\der m)$) is chosen to have a central galaxy brighter than the luminosity threshold $M_{r,{\rm max}}$. Each $m$-halo with a central is then assigned a number of satellites drawn from a Poisson distribution with mean $\bar N_{\rm sat}(<M_{r,{\rm max}}|m)$. The luminosities of centrals and satellites are then assigned using the distributions $f_{\rm cen}(<M_{r}|m)/f_{\rm cen}(<M_{r,{\rm max}}|m)$ and $\bar N_{\rm sat}(<M_{r}|m)/\bar N_{\rm sat}(<M_{r,{\rm max}}|m)$, respectively.
The functions $f_{\rm cen}(<M_{r}|m)$ and $\bar N_{\rm sat}(<M_{r}|m)$ define the HOD, with the mean number of galaxies brighter than $M_r$ residing in $m$-halos given by $\avg{N_{\rm gal}|m} = f_{\rm cen}(<M_{r}|m)\left[1+\bar N_{\rm sat}(<M_{r}|m)\right]$. In this work we will use the forms calibrated by \cite{zehavi+11}; specifically, we use an interpolation kindly provided by Ramin Skibba. 
Our fiducial cosmology has the same parameter values as used by \citet{zehavi+11} for their HOD analysis.
Satellite luminosities are assigned after the central ones, and the algorithm ensures that the central is the brightest galaxy of the group. We will return to this point later.

Each central is placed at the center-of-mass of its parent dark matter halo and is assigned the velocity of the halo.  The satellites are distributed around the centrals according to a truncated Navarro-Frenk-White (NFW) profile \citep{nfw96}, using a halo concentration as described below, and are assigned random velocities relative to the central that are drawn from a Maxwell-Boltzmann distribution that scales with halo mass\footnote{We note that our analysis in this paper does not use information regarding galaxy velocities; this will however be essential in future work when comparing, e.g., to observations of clustering in redshift space.}. This procedure for assigning satellite positions and velocities is more efficient, but perhaps less accurate, than identifying satellites with subhalo positions (e.g., there is no information regarding infall times and number of pericenter passages, but we need not run a very high resolution simulation and track low mass subhalos).


\subsubsection{Galaxy colours and the central-satellite split}
\noindent
Having assigned galaxy positions, velocities and luminosities, galaxy colours are assigned by drawing from  double-Gaussian fits to the observed distribution of $g-r$ colours at fixed luminosity. The two components of the double-Gaussian -- `red' and `blue' -- have means, variances and relative fraction as functions of luminosity that are fit to data. We give the results of such fits to SDSS data in section~\ref{sec:data:subsec:p(g-r|Mr)}.
Since these fits work at the level of the full galaxy sample, there is some freedom in deciding what fraction of centrals and satellites at fixed luminosity must be labelled `red' with the above definition. This is a somewhat subtle issue which merits some discussion.

In general, one might assume that the central and satellite red fractions depend on both luminosity of the object as well as the mass of the parent halo, and we would denote these quantities as $p({\rm red}|{\rm cen},M_r,m)$ and $p({\rm red}|{\rm sat},M_r,m)$, respectively. S09 showed that a model in which these quantities do not depend on halo mass (but have a non-trivial luminosity dependence) is consistent with measurements of colour-marked clustering of SDSS galaxies. However, colour-dependent clustering measurements will also depend on the level of conformity, which is what we are trying to model here (and which S09 did not do). We therefore adopt a simpler approach: we continue to assume that these red fractions are independent of halo mass, but fix the form of $p({\rm red}|{\rm sat},M_r)$ by demanding agreement with a direct measurement of this quantity in an SDSS-based group catalog \citep[][henceforth, Y07]{yang+07}. The form of $p({\rm red}|{\rm cen},M_r)$ is then fixed by the assumed HOD, while the \emph{all}-galaxy red fraction $p({\rm red}|M_r,m)$ inherits a mass dependence from the HOD (see Appendix~\ref{app:massdepredfrac})\footnote{See \cite{skibba09} for a comparison of central and satellite colours using mocks based on the S09 algorithm and in an earlier version of the Y07 catalog.}. In Appendix~\ref{app:massdepredfrac} we also explore the consequences of a halo mass dependence in the red fraction of centrals. 

The S09 algorithm has been used in several studies of galaxy environments \citep[e.g.,][]{muldrew+12,skibba+13} and the luminosity and color dependence of galaxy clustering \citep[e.g.,][]{skibba+14,carretero+15}.
We modify and extend this algorithm in two ways: (i) we correlate galaxy colours with the parent halo concentration and (ii) we calculate a stellar mass for each galaxy using a colour-dependent mass-to-light ratio.


\subsection{Correlating colour and concentration}
\label{sec:algo:subsec:ccc}
\noindent
As indicated above, in the basic S09 algorithm the colour of any given galaxy is assigned in two steps: 
first the galaxy is labelled `red' (a `red flag' is set to $1$) with probability $p({\rm red}|{\rm gal},M_r)$ where `gal' is either `cen' or `sat', and then its colour is drawn as a Gaussian random number with the appropriate mean and variance.
We correlate galaxy colours with parent halo concentrations by modifying the first step to adjust the red flag according to the concentration.
The trend we wish to introduce is that more concentrated, older halos should host older, or redder, galaxies.

At fixed halo mass, halo concentration $c$ is approximately Lognormally distributed, i.e. $\ln c$ is approximately Gaussian distributed with constant scatter $\sigma_{\ln c}\simeq 0.14\ln(10)$ and a mean value 
$\avg{\ln c} \equiv \ln\bar c(m,z)$ 
that depends on halo mass and redshift \citep[see, e.g.][]{bullock+01,wechsler+02}. In the range of interest the mean value is well described by \citep{ludlow+14}
\be
\bar c(m,z) = \alpha\, \nu(m,z)^{-\beta}\,,
\label{lognormal}
\ee
where $\nu(m,z)=\delta_{\rm c}(z)/\sigma(m)$ is the dimensionless ``peak height'' of the halo\footnote{$\delta_{\rm c}(z)$ and $\sigma(m)$ are, respectively, the critical density for spherical collapse at redshift $z$ and the r.m.s. of initial fluctuations smoothed on mass scale $m$, each extrapolated to $z=0$ using linear theory.} and the coefficients $\alpha$ and $\beta$ depend on the choice of halo mass definition. We will construct catalogs using the $m_{\rm 200b}$ definition for which the \citet{zehavi+11} HOD was calibrated ($m_{\rm 200b}$ is the mass contained in the radius $R_{\rm 200b}$ where the spherically averaged density is $200$ times the background matter density $\bar\rho(z)$) and the corresponding $\bar c(m,z)$ relation. For this choice we have $\alpha=7.7$ and $\beta=0.4$ \citep[see, e.g., Appendix A of][]{paranjape14}. 

Let $p({\rm red})$ be the value of the red fraction in the \emph{absence} of any correlation with concentration. (For clarity we drop the explicit dependence on galaxy luminosity and type.) It is useful to define the quantity 
\be
s\equiv\frac{\ln(c/\bar c)}{\sigma_{\ln c}}\,,
\label{s-def}
\ee
whose distribution $p(s)$ is Gaussian with zero mean and unit variance,
with $\bar c(m)$ and $\sigma_{\ln c}$ defined above.
We compute a conditional red fraction $p({\rm red}|s)$ that depends on the parent halo concentration by setting
\be
p({\rm red}|s) = (1-\rho)p({\rm red}) + \rho\,\Theta(s-s_{\rm red})\,,
\label{newredfrac}
\ee
where the Heaviside function $\Theta(x)$ is unity for $x>0$ and zero otherwise.
This says that galaxies in halos with concentrations $s > s_{\rm red}$ have an enhanced probability to be red as compared to the S09 model, while this probability is lowered by a factor $(1-\rho)$ compared to $p(\rm red)$ for galaxies in low-concentration halos. The dividing line $s_{\rm red}$ between high and low concentrations is defined in such a way that the \emph{average} red fraction across all halos satisfies
\be
\avg{p({\rm red}|s)} = \int_{-\infty}^\infty\der s\,p(s)\,p({\rm red}|s) = p({\rm red})\,,
\label{pred-avg}
\ee
which, using \eqn{newredfrac}, implies
\be
p({\rm red}) = p(s > s_{\rm red}) = \frac{\textrm{erfc}(s_{\rm red}/\sqrt{2})}{2}\,.
\label{sred-defn}
\ee
Equation~\eqref{newredfrac} gives a step-like dependence of the red fraction on concentration; in principle one could also imagine schemes where the red fraction increases with concentration in a continuous manner, but in this case it becomes more complicated to simultaneously ensure $\avg{p({\rm red}|s)} = p({\rm red})$ and $0 < p({\rm red}|s) < 1$, due to the Gaussian integral in \eqn{pred-avg}.

The parameter $\rho$ lies between zero and unity and controls the strength of the correlation between the red fraction and concentration; we discuss its physical meaning below. Setting $\rho = 0$ gives us the uncorrelated case when the red fraction does not depend on parent halo concentration. Setting $\rho = 1$ on the other hand corresponds to `complete correlation' where \emph{all} galaxies in high(low)-concentration halos are labelled red (blue). The intermediate case $0<\rho<1$ clearly interpolates between these two extremes.

Having set up our model, we proceed by assigning colours to the galaxies. The red flag of the galaxy is chosen by drawing a uniform random number $u\in[0,1)$ and setting the red flag to unity if $u < p({\rm red}|s)$ and zero otherwise. Next, we draw a Gaussian random number $g-r$ from the appropriate red or blue distribution. In the basic S09 algorithm, the latter step also does not depend on parent halo concentration. In principle, we could once again use the values of $s$ to introduce a further correlation between the actual $g-r$ values and halo concentration, but in practice it turns out to be difficult to do this while preserving the global colour distributions at fixed luminosity (which we must do since these are constrained by data). We therefore stick to the S09 procedure at this stage and draw from the appropriate `red'/`blue' Gaussian without further correlation with concentration. 

Ideally, we would correlate galaxy colours with the concentrations actually measured in the simulation. However, these concentrations are only approximately Lognormal, and deviations from the Lognormal shape render the second equality in \eqn{sred-defn} above invalid. Continuing to use $s_{\rm red} = \sqrt{2}\textrm{erfc}^{-1}(2p({\rm red}))$ then leads to the overall central and satellite red fractions not being preserved, with $\avg{p({\rm red}|s)} \neq p({\rm red})$. We could correct for this by calibrating the exact shape of $p(s)$ and numerically inverting the relation $p(s > s_{\rm red}) = p(\rm red)$ to obtain $s_{\rm red}$.
Note, however, that what we are really interested in is not the actual shape of the distribution of concentrations, but only their \emph{ranking} in halos of fixed mass. It is therefore easier to simply ``Gaussianize'' the measured log-concentrations, proceeding as follows. 
\begin{enumerate}
\item We first randomly draw a Gaussian variate $s$ for each halo in the catalog and derive a Lognormal concentration $c$ using \eqns{lognormal} and~\eqref{s-def}. 
\item Next, we bin the halos in $16$ equi-log-spaced mass bins\footnote{The number of bins was chosen after testing for convergence.} and consider the lists of measured and Lognormal concentrations of halos in each bin. We rank order and reassign the Lognormal concentrations in a given bin according to the measured concentrations in that bin. In detail, if there are $N$ halos in a bin, then the halo with the $j^{\rm th}$ largest measured concentration is assigned the $j^{\rm th}$ largest Lognormal concentration in this bin, with $j=1,2,\ldots,N$.
\item These reassigned Lognormal values of $c$ are then used to distribute satellites in their respective parent halos, and the corresponding Gaussian values of $s$ are used to induce conformity in the colours as described above, with $s_{\rm red}$ given by \eqn{sred-defn}. 
\end{enumerate}
This ensures that galaxy colours in this model are assigned consistently and inherit the large scale environment dependence of assembly bias.

We will show that this procedure, which defines our default model, leads to differences in the satellite red fraction in groups with blue and red centrals (1-halo conformity) as well as specific spatial trends of galaxy red fractions around blue and red isolated objects well outside $R_{\rm 200b}$ (2-halo conformity). 
As mentioned in the Introduction, it is also useful to ask how these trends would change if conformity arises not because of assembly bias, but due to some other halo property that \emph{does not} show environmental trends. 
To this end, we will also consider a model in which galaxy colours are chosen and satellites are distributed exactly as in the default procedure above, except that we skip the rank ordering step (ii). This is equivalent to randomizing the halo concentrations at fixed halo mass before correlating them with the galaxy colours.
In this model -- which we will denote ``\emph{no-2h}'' --  we expect our mocks to exhibit 1-halo conformity without a corresponding 2-halo signal.

The parameter $\rho$ has an interesting interpretation. It is easy to show that $\rho$ satisfies the relations
\begin{align}
\rho &=\, \frac{p({\rm red}|s>s_{\rm red}) - p({\rm red})}{1 - p({\rm red})}\notag\\
&\sim\, \frac{p({\rm red}|{\rm old}) - p({\rm red})}{1 - p({\rm red})}\,.
\label{understanding-rho}
\end{align}
where the second approximation holds if we assume that high concentration halos are old.
Although simply a restatement of the definition of $\rho$ in \eqn{newredfrac}, \eqn{understanding-rho} allows us to interpret $\rho$ as a ``quenching efficiency'' \citep{vdb+08,peng+10,peng+12,kovac+14,phillips+14,klwk14}. Written like this, in our default model $\rho$ is the fraction of blue (or star forming) galaxies that became red (or quenched) because their respective parent halos grew old\footnote{This can only be approximately correct, since, e.g., the present day population of blue satellites is not in general representative of the progenitors of all present day satellites. Our results do not depend on this interpretation, however, and one could simply treat $\rho$ as a free parameter in the model.}. 
In the \emph{no-2h} model, however, the connection between halo ages and galaxy colours is lost (since the former exhibit halo assembly bias at large scales while the latter will not), and the second relation in \eqn{understanding-rho} will not hold. 
In each model of conformity, however, $\rho$ can be thought of as a ``group quenching efficiency'', driven by the age of the parent halo of the group in the default model and by some local but otherwise unspecified property of the group in the \emph{no-2h} model.

Our scheme of generating concentration-dependent red flags smoothly interpolates between the uncorrelated case ($\rho=0$) and the nearly `completely correlated' case ($\rho=1$). In practice, we perform this operation separately for satellites and centrals, which ensures that the overall central and satellite red fractions are left unaltered by construction. The model is flexible enough that, if needed, the group quenching efficiency $\rho$ can be set separately for centrals and satellites, with the interpretation that different quenching mechanisms might play a role for centrals and satellites in the same halo. E.g., this might be the case if \emph{sub}halo age is more relevant for satellite colour than the age of the parent halo \citep[][]{hw13}. In this work, however, we do not explore this possibility and only use a one-parameter setup with $\rho$ taken to be the same for centrals and satellites, independent of galaxy luminosity and halo mass. Galactic conformity arises from the fact that the colours of centrals and satellites in a group are affected by their common group concentration. The fact that the concentrations have a scatter at fixed halo mass means that the conformity will persist even when binning objects by the parent halo mass.


\subsection{Calculating stellar masses}
\label{sec:algo:subsec:sm}
\noindent
As a separate step, having assigned galaxy colours, we compute stellar masses using a colour dependent mass-to-light ratio that we have also fit to SDSS galaxies. In particular, we use
\be
(M/L)_r = 4.66(g-r)-1.36(g-r)^2-1.108\,,
\label{masstolight-fit}
\ee
to compute the ${}^{0.1}r$-band Petrosian mass-to-light ratio $(M/L)_r$ in units of $\Msun/L_{\odot}$, and add a Gaussian $1$-sigma scatter of $0.2$ (we describe the procedure and data set used to obtain this fit in section~\ref{sec:data}). Taking the base-10 logarithm and adding $(M_r-4.76)/(-2.5)$ gives us the log stellar mass in units of $\Mhsq$. This does not guarantee that the central of a group, which is the brightest by construction, is also the most massive. In practice this affects about $8$-$10\%$ of the groups in the total sample, with the actual number depending on the the value of $\rho$. This may be related to the findings of \citet{skibba+11} regarding a non-zero fraction of groups in which the central is not the brightest; we will explore this further in future work.


\section{Simulations}
\label{sec:simulations}
\noindent
Our mocks are built on a suite of $N$-body simulations of cold dark matter (CDM), each of which resolves a periodic cubic box of comoving volume $(200\Mpch)^3$ with $512^3$ particles using our fiducial cosmology. This gives us a particle mass of $m_{\rm part}=4.1\times10^{9}\Mh$. The simulations were run using the tree-PM code\footnote{http://www.mpa-garching.mpg.de/gadget/} \textsc{Gadget-2} \citep{springel:2005} with a force resolution $\epsilon = 12.5h^{-1}$kpc comoving ($\sim1/30$ of the mean particle separation) and a $1024^3$ PM grid. 
Initial conditions were generated at $z=99$ employing $2^{\rm nd}$-order Lagrangian Perturbation Theory \citep{scoccimarro98}, using the code\footnote{http://www.phys.ethz.ch/$\sim$hahn/MUSIC/} \textsc{Music} \citep{hahn11-music} with a transfer function calculated using the prescription of \citet{eh98}. 
We have run $10$ realisations of this simulation by changing the random number seed used for generating the initial conditions and have used the resulting $z=0$ snapshots for our mocks. The simulations were run on the Brutus cluster\footnote{http://www.cluster.ethz.ch/index\_EN} at ETH Z\"urich.

To identify halos, we have used the code\footnote{http://code.google.com/p/rockstar/} \textsc{Rockstar} \citep{behroozi13-rockstar}, which assigns particles to halos based on an adaptive hierarchical Friends-of-Friends algorithm in $6$-dimensional phase space. \textsc{Rockstar} has been shown to be robust for a variety of diagnostics such as density profiles, velocity dispersions, merger histories and the halo mass function. As mentioned earlier, we use the $m_{\rm 200b}$ value reported by \textsc{Rockstar} as the mass of the parent halo, and $R_{\rm 200b}$ as its radius. 
We use the value of the halo scale radius $r_{\rm s}$ reported by \textsc{Rockstar} to compute halo concentration $c_{\rm 200b} = R_{\rm 200b}/r_{\rm s}$.
The smallest halo mass we resolve sets the faintest luminosity we can reliably sample. We discard objects having $m_{\rm 200b} < 20\,m_{\rm part}$, which allows us to set the luminosity threshold to $M_{r,{\rm max}}=-19.0$. 
Having $10$ independent simulations means that we can reliably characterise the sample variance on our observables for each mock configuration.


\section{Data}
\label{sec:data}
\noindent
In this section we describe the data sets we use for comparison with the mocks, and give details of various fits used in defining the mocks.


\subsection{The Y07 catalog}
\label{sec:data:subsec:Y07cat}
\noindent
We base some of the fits required for generating our mocks (the colour distribution and satellite red fraction at fixed luminosity, and the mass-to-light ratio at fixed colour) on the galaxies contained in the Y07 group catalog\footnote{http://gax.shao.ac.cn/data/Group.html}. These fits do not depend on the level of conformity, and we therefore additionally use the Y07 catalog as a baseline for galactic conformity as well. The Y07 catalog is built using the halo-based group finder described in \cite{yang+05} to identify groups in the New York University Value Added Galaxy Catalog \citep[NYU-VAGC;][]{blanton+05}, based on the SDSS\footnote{http://www.sdss.org} \citep{york+00} data release 7 \citep[DR7;][]{abazajian+09}. Throughout, we will only consider galaxies with spectroscopic redshifts (`sample II' of Y07) restricted to the range $0.01 < z < 0.07$, and also restrict the Petrosian absolute magnitudes to the range $-23.7 < M_r < -16$.

In part of the analysis below, we will use a luminosity-complete subsample -- denoted ``Y07-Lum'' --  containing galaxies with Petrosian absolute magnitudes $M_r < M_{r,{\rm max}} = -19.0$ and Model $g-r$ colours from the NYU-VAGC as provided by Y07. Additionally, when comparing properties as a function of stellar mass, we will use another subsample -- denoted ``Y07-Mass'' -- containing values of Petrosian stellar mass $m_\star$ and Model $g-r$ colours obtained by running the \textsc{kcorrect\_v4.2} code\footnote{http://howdy.physics.nyu.edu/index.php/Kcorrect} of \cite{br07} on the corresponding Petrosian and Model properties, respectively, of the Y07 galaxies obtained from the MPA-JHU catalog\footnote{http://www.mpa-garching.mpg.de/SDSS/DR7/} \citep{kauffmann+03} by matching galaxy IDs across the MPA-JHU and NYU catalogs. We use magnitudes that are K-corrected to their values at $z=0.1$. We have checked that there is reasonable agreement between comparable galaxy properties in the NYU-VAGC and the MPA-JHU catalog. When showing mass functions and fractions, we apply inverse $V_{\rm max}$-weighting to the galaxies in the Y07-Mass catalog.

\begin{figure}
\centering
\includegraphics[width=0.475\textwidth]{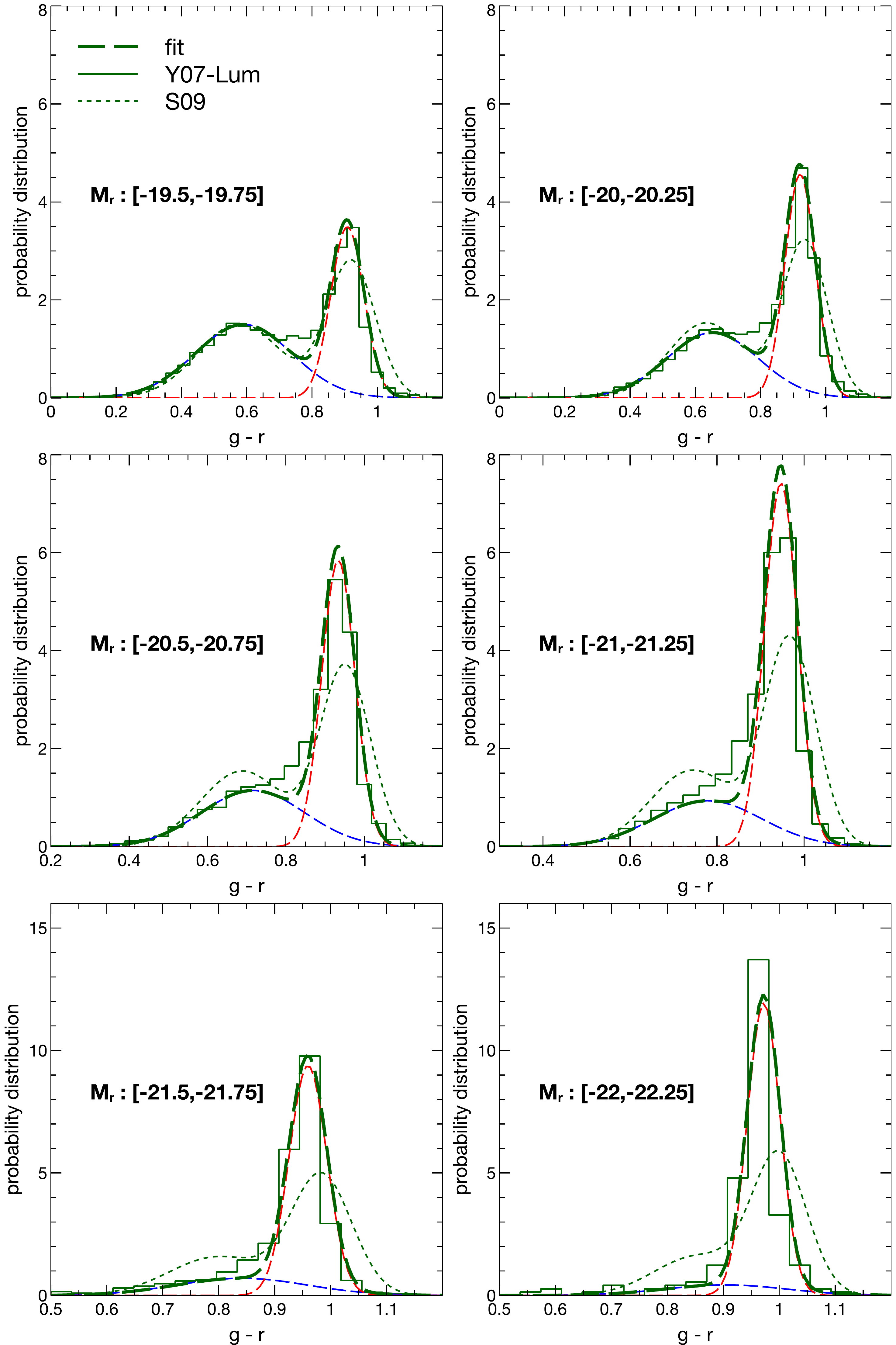}
\caption{Comparison of observed Model $g-r$ distributions in bins of Petrosian $M_r$ in the Y07-Lum catalog (solid green histograms) with the corresponding double-Gaussian fits reported in \eqns{dbl-Gauss-fits}. We show the `red' and `blue' components in the fits as the respectively red and blue dashed lines, and the total of these as the dashed green lines. For comparison we show the S09 fits as dotted lines; the discrepancy between S09 and the Y07 catalog is most likely due to differences between DR7 and DR4 (which was used by S09) and our choice of using Model colours.}
\label{fig:gr-comparemock}
\end{figure}

Our choice of Petrosian quantities for defining both the stellar masses as well as the mass-to-light fit (see below) in the Y07-Mass catalog is motivated by the fact that the HOD we use has been calibrated on Petrosian absolute magnitudes by \citet{zehavi+11}. We have checked that there is reasonable agreement between Model and Petrosian stellar masses in the Y07-Mass catalog and, moreover, that the trends in the red fractions and associated observables which we discuss below differ by only $\sim10\%$ when using Model or Petrosian quantities. Consequently, we do not expect any of our qualitative conclusions to be affected by our choice.

\begin{figure*}
\centering
\includegraphics[width=0.85\textwidth]{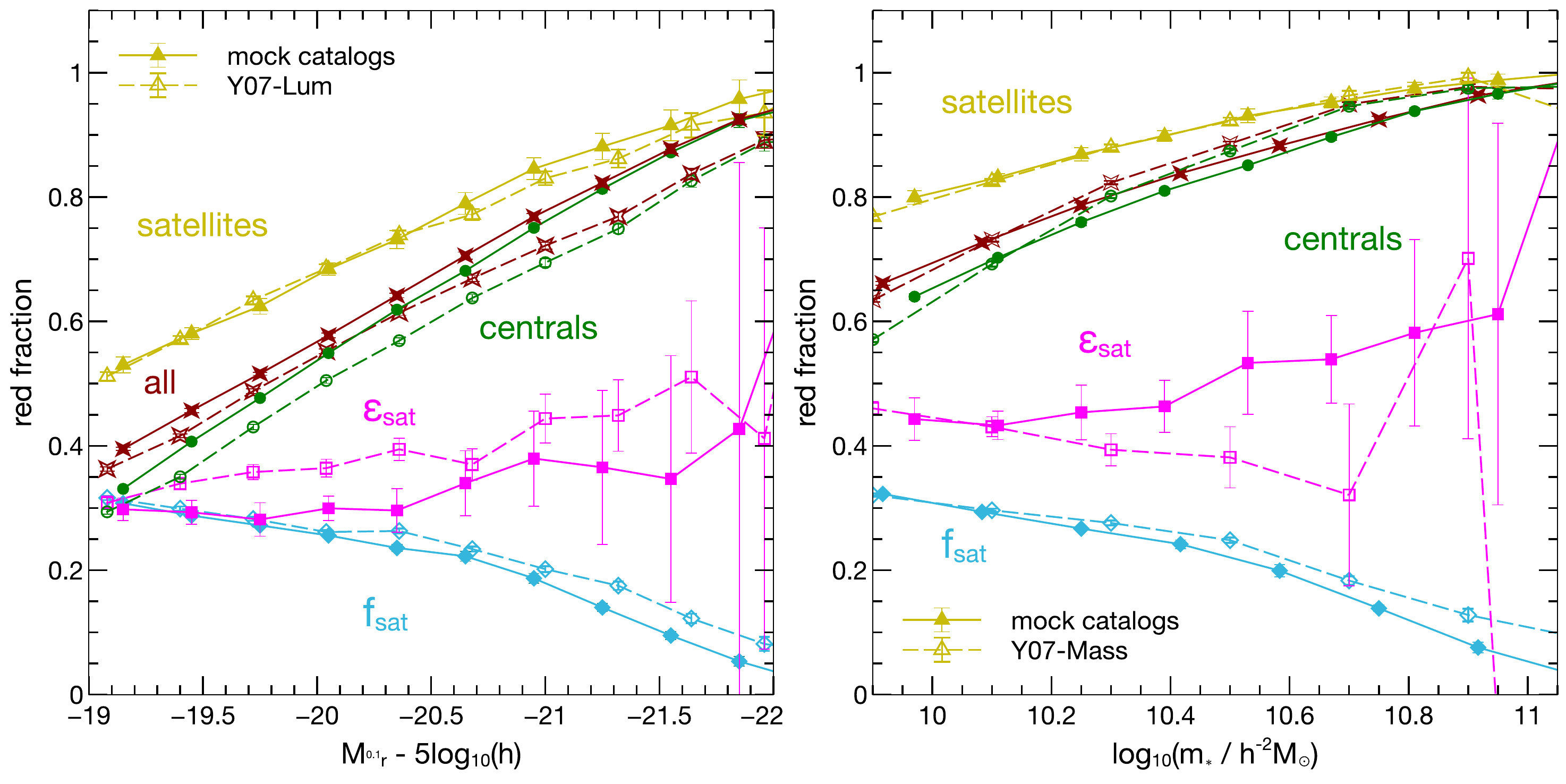}
\caption{Conformity-independent variables as a function of luminosity \emph{(left panel)} and stellar mass \emph{(right panel)}. Filled symbols joined by solid lines show the results of averaging over $10$ realisations of our mocks, while open symbols joined by dashed lines show measurements in the Y07 catalog. The error bars on the mock results show the r.m.s. fluctuations (standard deviation) around the mean value, while the errors on the Y07 data are estimated from 150 bootstrap resamplings. The cyan diamonds show the satellite fraction $f_{\rm sat}$. The dark red crosses, green circles and yellow triangles respectively show the red fractions for all galaxies ($f_{\rm red}$), only centrals ($f_{\rm red|cen}$) and only satellites ($f_{\rm red|sat}$). The magenta squares show the satellite quenching efficiency $\varepsilon_{\rm sat} = (f_{\rm red|sat}-f_{\rm red|cen})/(1-f_{\rm red|cen})$. The `red'/`blue' classification for both data and mocks uses \eqn{colorcut-Mr} for the left panel and \eqn{colorcut-sm} for the right panel.}
\label{fig:redfrac-avg}
\end{figure*}


\subsection{Mass-to-light ratio fit}
\label{sec:data:subsec:M/L}
\noindent
We fit a quadratic relation to the ${}^{0.1}r$-band Petrosian mass-to-light ratio $(M/L)_r$ at fixed value of Model $g-r$ colour in the Y07-Mass catalog. In particular, we compute the median value of $(M/L)_r$ in bins of $g-r$, using the median $g-r$ as the bin center.
The resulting least-squares best fit is given in \eqn{masstolight-fit} and is close to but slightly lower than the one in equation (1) from \cite{ww12}. The measured r.m.s. scatter of $(M/L)_r$ in the bins of $g-r$ has an average value of $0.2$ which we also use as described below \eqn{masstolight-fit}. 


\subsection{Color-luminosity fits}
\label{sec:data:subsec:p(g-r|Mr)}
\noindent
We fit double-Gaussian shapes to the distributions of Model $g-r$ colours in bins of Petrosian absolute magnitude $M_r$ in the Y07-Lum catalog. The resulting fits can be summarised using $5$ quantities: the means and variances of the `red' and `blue' distributions and the probability $p({\rm red}|M_r)$ that the colour is drawn from the `red' distribution. The full $g-r$ distribution can then be written as
\begin{align}
p(g-r|M_r) &= p({\rm red}|M_r)\,p_{\rm red}(g-r|M_r)\notag\\ 
&\ph{p({\rm red})}
+ \left(1-p({\rm red}|M_r)\right)\,p_{\rm blue}(g-r|M_r)\,,
\label{p-gr-full}
\end{align}
where, e.g., $p_{\rm red}(g-r|M_r)$ is Gaussian with mean $\avg{g-r|M_r}_{\rm red}$ and standard deviation $\sigma_{\rm red}(M_r)$, and similarly for the blue distribution. We find the following best fit values:
\begin{align}
p({\rm red}|M_r) &= 0.423 - 0.175\left(M_r+19.5\right) \notag\\
\avg{g-r|M_r}_{\rm red} &= 0.9050 - 0.0257\left(M_r+19.5\right) \notag\\
\avg{g-r|M_r}_{\rm blue} &= 0.575 - 0.126\left(M_r+19.5\right) \notag\\
\sigma_{\rm red}(M_r) &= 0.0519 + 0.0085\left(M_r+19.5\right) \notag\\
\sigma_{\rm blue}(M_r) &= 0.150 + 0.015\left(M_r+19.5\right)
\label{dbl-Gauss-fits}
\end{align}
Figure~\ref{fig:gr-comparemock} compares these analytic functions (dashed lines) with the measured $g-r$ distribution in the Y07-Lum catalog (solid histograms). 
For comparison, the dotted lines show the fits reported by S09 which were based on a similar catalog using DR4 of SDSS; these are rather different from the Y07-Lum data, which could be due to differences between the DR7 colours and those in DR4. (The discrepancy reduces but does not disappear if we use Petrosian colours.)
Overall, we see a good agreement between the data and the fits, except for a ``green valley'' which is not captured by the sum of two Gaussians. In principle, one can obtain a better fit by introducing a third `green' component \citep{krause+13,carretero+15}, but we choose not to do this here. 

\begin{figure*}
\centering
\includegraphics[width=0.85\textwidth]{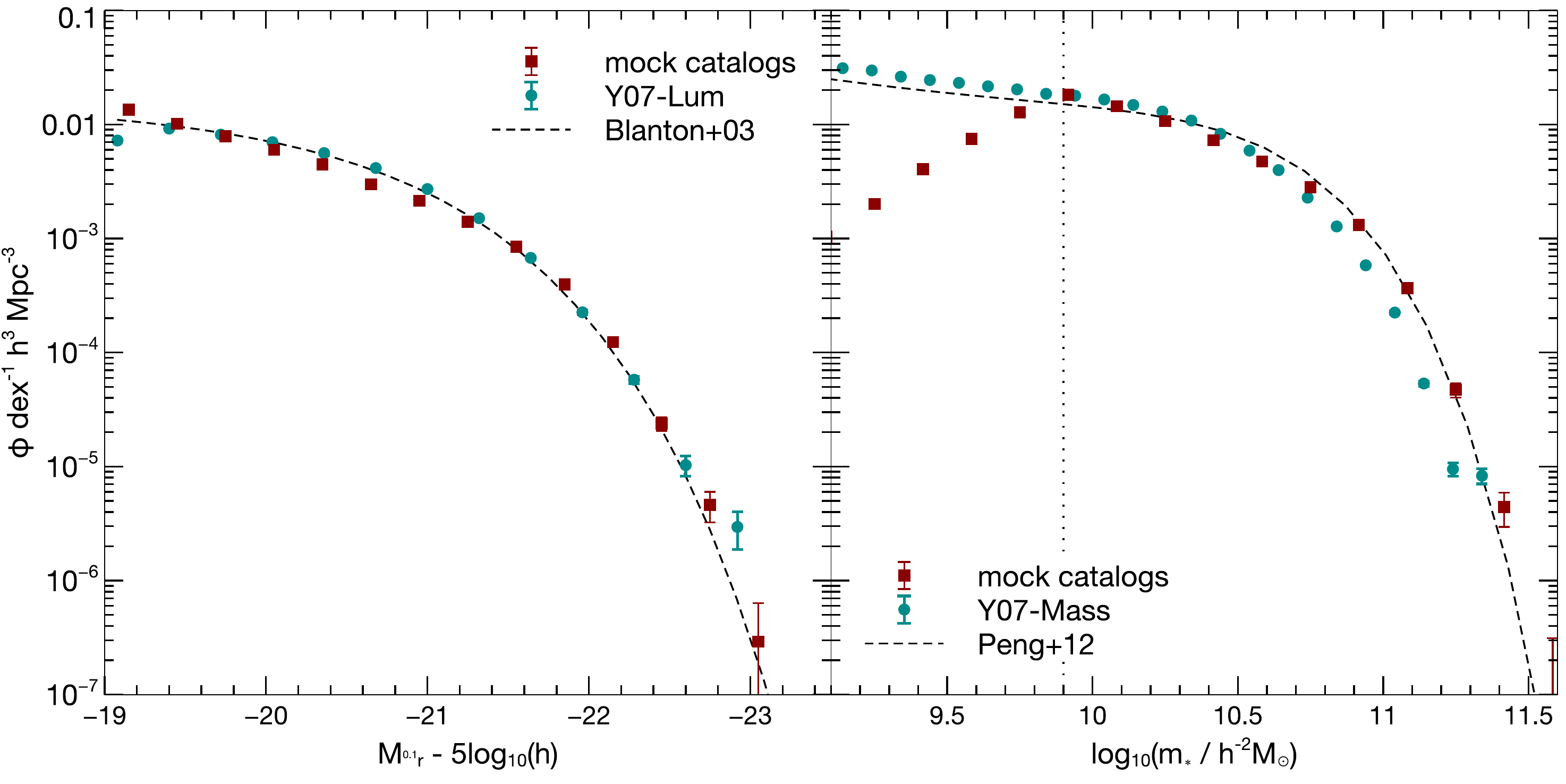}
\caption{All-galaxy luminosity \emph{(left panel)} and stellar mass functions \emph{(right panel)}. 
Red squares show the average over $10$ independent mock catalogs, with error bars showing the r.m.s. scatter. 
Cyan circles show measurements in the Y07-Lum (left) and Y07-Mass catalogs (right), with errors estimated from 150 bootstrap resamplings. 
For comparison, the short-dashed black curve in the left panel shows the Schechter fit to the ${}^{0.1}r$-band luminosity function in SDSS data reported by \citet{blanton+03}, while the corresponding curve in the right panel shows the stellar mass function fit for the Y07 catalog as reported by \citet{peng+12}. 
The vertical dotted line in the right panel shows the approximate mass-completeness limit of our mocks at $\log_{10}(m_\star)=9.9$.}
\label{fig:massfuncs}
\end{figure*}

Although the algorithm classifies objects as red or blue depending on which Gaussian distribution their colours are drawn from, for ease of comparison with observational results, we will use sharp thresholds in $g-r$. When showing results as a function of luminosity we classify objects as red when their $g-r$ colour exceeds
\be
(g-r)_{\rm cut} = 0.8 - 0.03\left(M_r+20\right)\,,
\label{colorcut-Mr}
\ee
and as blue otherwise \citep{zehavi+05}. When showing results as a function of stellar mass we instead use the  threshold
\be
(g-r)_{\rm cut} = 0.76 + 0.10\left[\,\log_{10}(m_{\star})-10\,\right]\,,
\label{colorcut-sm}
\ee
which is somewhat shallower than the threshold quoted by \citet{vdb+08} who use a slope of $0.15$ instead of $0.1$. We have found that \eqn{colorcut-sm} gives a slightly better separation of the bi-modal colour-mass distribution in the Y07-Mass catalog than the \citet{vdb+08} relation does, although our qualitative conclusions do not depend on the choice of threshold. We apply the same threshold to colours in our mocks as well as in the Y07 data. Using sharp thresholds means that the measured red fractions $f_{\rm red}$ will, in general, differ from the probabilities $p({\rm red})$ discussed earlier.


\section{Results}
\label{sec:results}
\noindent
We now present the results of our mock algorithm and compare them with corresponding measurements in the Y07 catalog. We start by discussing observables that \emph{do not} depend on the presence or absence of conformity.


\subsection{Conformity-independent observables}
\label{sec:results:subsec:noconf}
\noindent
The following observables are independent of the level of conformity (i.e., the value of $\rho$) because their definition does not involve a simultaneous determination of central and satellite colour:
the fraction of galaxies that are satellites ($f_{\rm sat}$), the average red fractions of all galaxies, centrals and satellites (respectively, $f_{\rm red}$, $f_{\rm red|cen}$ and $f_{\rm red|sat}$), and the satellite quenching efficiency $\varepsilon_{\rm sat} = (f_{\rm red|sat}-f_{\rm red|cen})/(1-f_{\rm red|cen})$. 
Among these functions, the luminosity dependence of $f_{\rm sat}$ is determined solely by the HOD (which itself is fit using the luminosity dependence of clustering), that of $f_{\rm red}$ is fixed by the double-Gaussian fits\footnote{Note that $f_{\rm red}$ depends on the actual shape of the double-Gaussian, not only on $p({\rm red}|M_r)$.}, and the remaining quantities additionally depend on the choice of satellite red fraction which we discuss next. 

As mentioned previously, we restrict ourselves to a simple prescription in which $p({\rm red}|{\rm sat},M_r)$ (the probability that $g-r$ for a satellite of luminosity $M_r$ is drawn from the `red' Gaussian) \emph{as well as} the corresponding quantity $p({\rm red}|{\rm cen},M_r)$ for centrals are independent of halo mass, in which case only one of these can be chosen independently. By comparing to the measured luminosity dependence of $f_{\rm red|sat}$, we find reasonable agreement using
\be
p({\rm red}|{\rm sat},M_r) = 1.0 - 0.33\left[1+\tanh\left(\frac{M_r+20.25}{2.1}\right)\right]\,,
\label{p-redsat}
\ee
although this is by no means the only function that can do the job. This in turn fixes the luminosity dependece of $f_{\rm red|cen}$ and $\varepsilon_{\rm sat}$. The stellar mass dependence of all the quantities mentioned above is then also completely fixed. 

The results of the comparison are shown in Figure~\ref{fig:redfrac-avg}, as a function of luminosity in the left panel and stellar mass in the right panel. The open symbols joined by dashed lines show the measurements in the Y07-Lum (left panel) and Y07-Mass catalogs (right panel), while the filled symbols joined by solid lines show the mean over 10 realisations of our mocks. The error bars on the mock data show the r.m.s. scatter (sample variance) of the respective quantities over all realisations, while the error bars on the Y07 data are estimated from 150 bootstrap resamplings. 
We see that, apart from systematic differences of $\lesssim10\%$, there is good agreement between the mocks and the data. In particular, the mocks correctly reproduce the observed near-independence of $\varepsilon_{\rm sat}$ on stellar mass and luminosity, as well as matching its amplitude. The systematic differences could be due to one or more of the following: residual systematics in the HOD parameters and the mass-to-light fit, the fact that we do not exactly reproduce the observed colour distribution, and our choice of halo mass-independent central and satellite red fractions. In particular, the difference in $f_{\rm red}$ is very likely due to the mismatch in the colour distributions around the ``green valley'' (see Section~\ref{sec:data:subsec:p(g-r|Mr)} and Figure~\ref{fig:gr-comparemock}). Since we match the satellite red fraction $f_{\rm red|sat}$ quite well by construction, the difference in $f_{\rm red}$ then also causes a difference in $f_{\rm red|cen}$.

\begin{figure}
\centering
\includegraphics[width=0.45\textwidth]{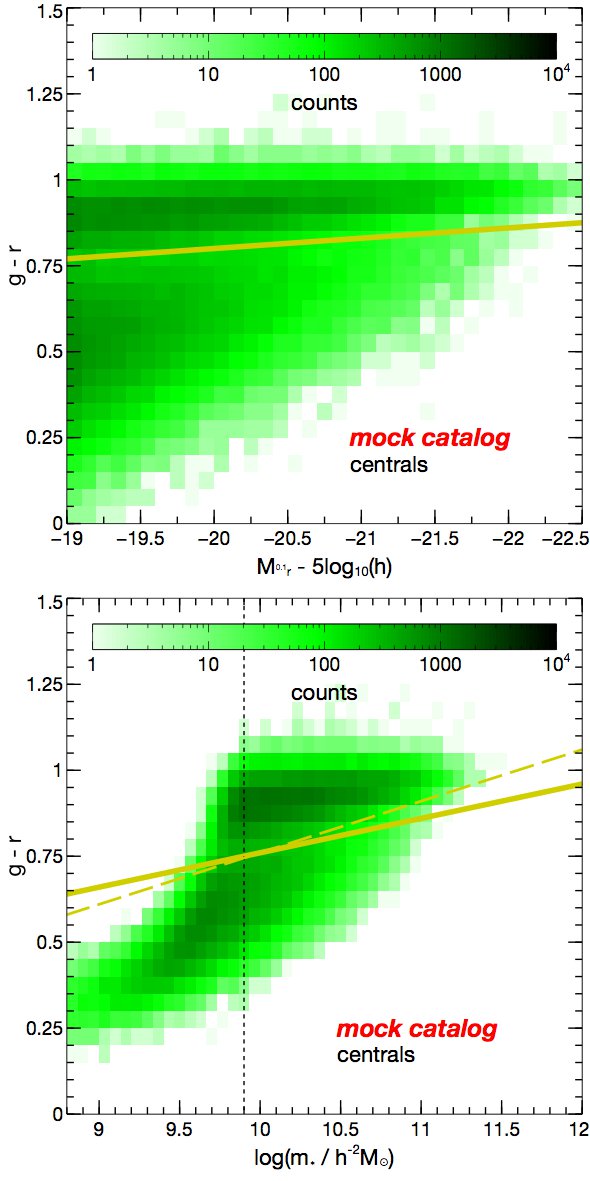}
\caption{Number counts of centrals as a function of $g-r$ and luminosity \emph{(top panel)} or stellar mass \emph{(bottom panel)} in one of our mocks. We clearly see how a luminosity threshold propagates into a mass incompleteness for red objects. A visual inspection indicates that our mocks should be mass complete for $\log_{10}(m_\star) \gtrsim 9.9$ (vertical dotted line in the bottom panel). The thick yellow lines show \eqn{colorcut-Mr} in the top panel and \eqn{colorcut-sm} in the bottom panel. For comparison, the dashed yellow line in the bottom panel shows the threshold used by \citet{vdb+08}.}
\label{fig:hist2d-cen}
\end{figure}

Finally, an important set of observables that are independent of conformity are the luminosity and stellar mass functions of the mocks. Figure~\ref{fig:massfuncs} compares these with corresponding measurements in, respectively,  the Y07-Lum and Y07-Mass catalogs. We also show the fits to SDSS data reported by \cite{blanton+03} for the luminosity function and \cite{peng+12} for the mass function.
Overall there is good agreement between the measurements in our mocks, in the data and the corresponding fits from previous work; this serves as a sanity check on our algoritm. The fact that our measurements in the Y07-Mass catalog do not quite agree with the fits reported by \citet{peng+12} highlights the fact that the shape and amplitude of the mass function is quite sensitive to the exact definition of stellar mass. 

From a visual inspection, the mocks appear to be mass complete for $\log_{10}(m_\star) \gtrsim 9.9$, which is the threshold we will use below. For our choice of $h=0.7$ in the $N$-body simulations, this corresponds to $\log_{10}(m_\star/\Msun) > 10.2$. The incompleteness at lower masses is because our mock is luminosity complete with $M_r < -19.0$; the colour-dependence of the mass-to-light ratio \eqref{masstolight-fit} then means that we are missing most of the faint red galaxies (and many of the faint blue ones) that would populate the low mass end. This becomes clearer in Figure~\ref{fig:hist2d-cen} which shows number counts of centrals as a function of $g-r$ and luminosity (top panel) or stellar mass (bottom panel) in one of our mocks. The deficit of low-mass red objects is clear in the bottom panel. Similar results hold for the satellites as well. Our choice of mass-completeness threshold is further justified by the fact that only $\sim1\%$ of the galaxies with $\log_{10}(m_\star)>9.9$ belong to the faintest bin $-19.2<M_r<-19.0$, implying that fainter galaxies would contribute negligibly above this mass threshold.


\subsection{Effects of correlating galaxy colour and halo concentration}
\label{sec:results:subsec:conf}
\noindent
We now investigate the effects of a non-zero value of $\rho$ in observables that \emph{do} respond to a correlation between central and satellite colours. To start with we simply explore the effects of changing the value of $\rho$, and later motivate a specific value by comparing to the Y07 catalog.

\begin{figure*}
\centering
\includegraphics[width=0.85\textwidth]{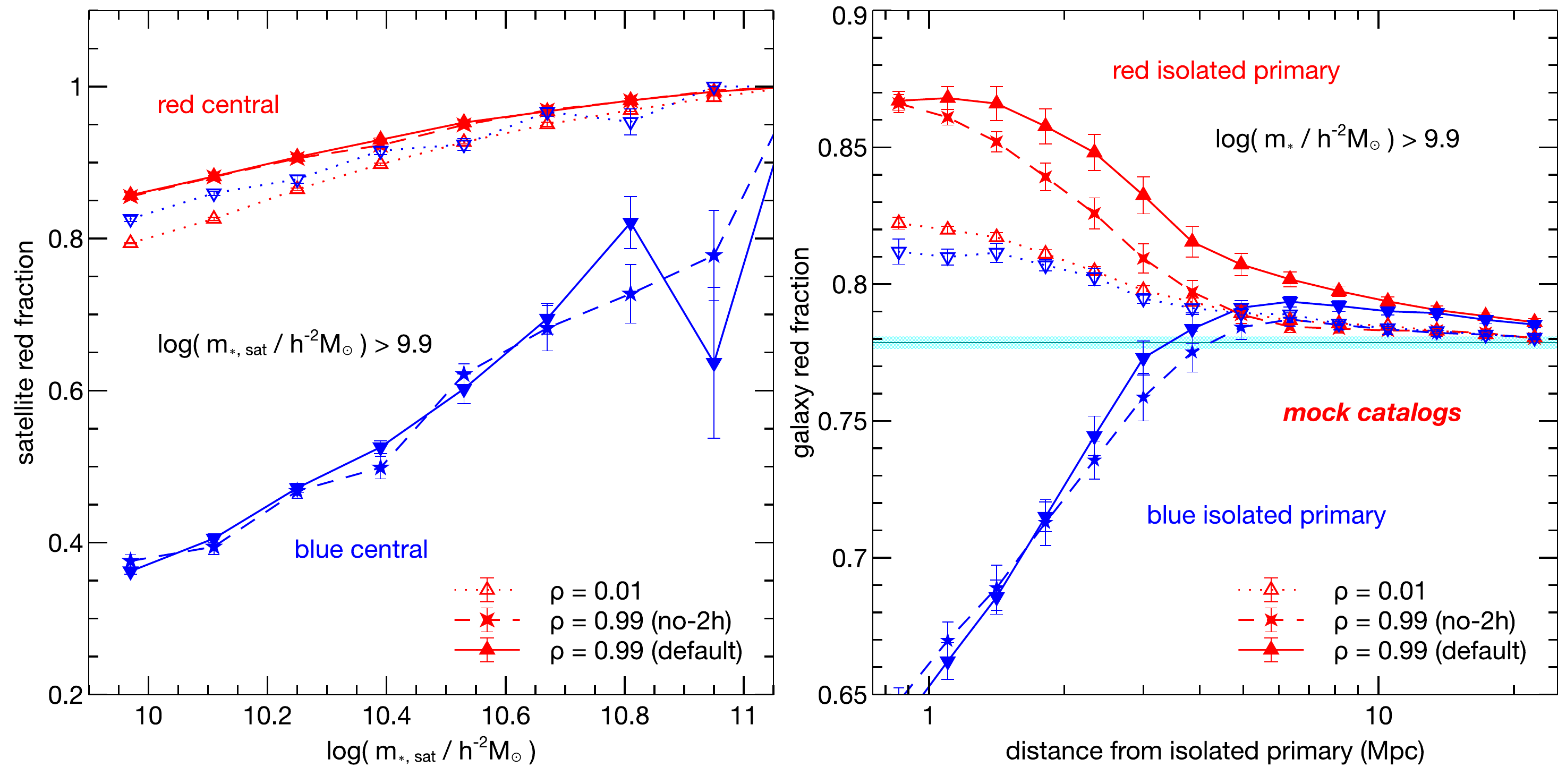}
\caption{\emph{(Left panel): }Satellite red fraction as a function of satellite stellar mass, split according to whether the associated centrals are red (red points and lines) or blue (blue points and lines). 
The open red triangles and blue inverted triangles with dotted lines show the result of the algorithm setting $\rho=0.01$, i.e. essentially \emph{no correlation} between galaxy colour and host halo concentration, as is routinely assumed by standard HOD based algorithms. 
The filled symbols with solid/dashed lines show the result when $\rho=0.99$, i.e. \emph{strong correlation}, in the presence (default model; red triangles and blue inverted triangles with solid lines) or absence (\emph{no-2h} model; red crosses and blue stars with dashed lines) of 2-halo conformity as described in the text.  
The points show the mean over 10 independent mocks and the error bars the standard error on the mean. 
\emph{(Right panel): }The red fraction of all galaxies surrounding red or blue ``isolated primaries'' as defined in the text, as a function of spherical distance, colour-coded and formatted as in the left panel, for the same set of 10 mocks used for the respective measurements in the left panel. 
All objects are required to have $\log_{10}(m_\star) > 9.9$ and we do not restrict the masses of their parent halos.
The horizontal cyan line and band respectively show the average all-galaxy red fraction above this stellar mass threshold and its r.m.s. scatter over 10 independent mocks (these values are independent of $\rho$).
Note that the scale on the vertical axis differs from that in the left panel. 
For both centrals and satellites, the distinction between red and blue was made using \eqn{colorcut-sm}.}
\label{fig:redfrac-2hconf}
\end{figure*}

\begin{figure*}
\centering
\includegraphics[width=0.85\textwidth]{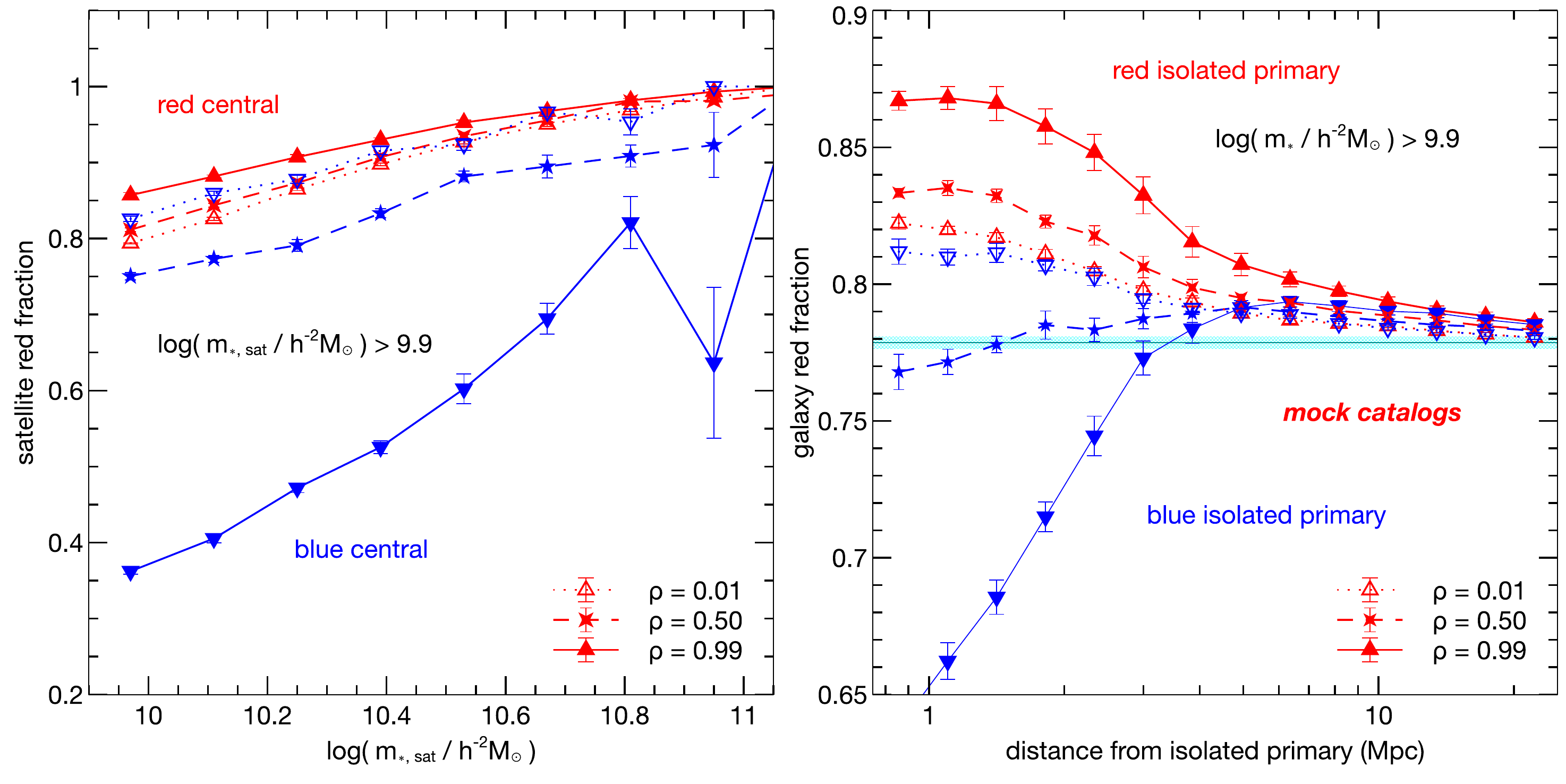}
\caption{Same as Figure~\ref{fig:redfrac-2hconf}, except that the filled red crosses and blue stars joined by dashed lines now show the result for mocks with the default model, but where we set $\rho=0.5$ (``medium'' correlation between colour and concentration). The open triangles with dotted lines and filled triangles with solid lines are the same as in Figure~\ref{fig:redfrac-2hconf}. This plot demonstrates the tunability of our model, which is one of our key results.}
\label{fig:redfrac-cc}
\end{figure*}

The left panel of Figure~\ref{fig:redfrac-2hconf} shows $f_{\rm red|sat}$ as a function of satellite stellar mass, split by whether the corresponding central is red or blue, for two choices of $\rho=0.01,0.99$. For each of these choices, the points show the mean over 10 independent mocks. Since we are interested in understanding these mean trends in this plot, the error bars in both panels show the standard error on the mean (rather than the r.m.s. scatter). 
The plot clearly shows that a positive correlation goes in the direction of explaining conformity, which has been noted earlier by others as well. The open symbols (red triangles and blue inverted triangles) joined by dotted lines show the result of the algorithm setting $\rho=0.01$, i.e. essentially no correlation, in keeping with what is usually assumed by standard HOD based algorithms. In this case there is no distinction between the two red fractions at the bright end, but fainter satellites with blue centrals are preferentially slightly redder than those with red centrals (a weak `negative conformity'). We have checked that the latter effect is due to averaging over luminosities and completely disappears if we plot results at fixed luminosity instead\footnote{This traces back to our choice of satellite red fraction which is independent of halo mass. Had this not been the case, we would have seen a similar effect in the luminosity-based plot as well, arising in this case from an averaging over halo mass. The effect in the stellar mass plot would now be even more pronounced.} (not shown).  

The filled symbols with solid/dashed lines show the result when $\rho=0.99$, i.e. strong correlation. In this case satellites with red centrals are clearly significantly redder than satellites with blue centrals.
The filled red triangles and blue inverted triangles joined by solid lines show the result for mocks that used our default model of conformity, rank ordering the Lognormal halo concentrations according to the measured concentrations from the simulation in $16$ equi-log-spaced mass bins (see Section~\ref{sec:algo:subsec:ccc}).
The red crosses and blue stars with dashed lines show the result for mocks that used the \emph{no-2h} model, with randomly assigned Lognormal concentrations.
As expected, the red fractions for these two cases are nearly identical. 

The right panel of the Figure is formatted identically to the left panel and uses the same set of mocks. In this case, we show the red fraction of all galaxies with $\log_{10}(m_\star)>9.9$ as a function of their distance $r$ from red or blue ``isolated primaries''; we define the latter as galaxies of mass $m_{\star}$ that do not have any galaxy more massive than $m_{\star}/2$ within a spherical radius of $500$kpc\footnote{This definition is somewhat different from the observationally-motivated one employed by \cite{kauffmann+13}, who used a cylinder in redshift space and projected distance \citep[see also][]{hwv15}. In this paper, however, we do not compare our 2-halo results with observations, and the simpler 3-dimensional definition above suffices to understand various trends in the signal. The signal itself is expected to diminish in strength due to projection effects.}.
We find that the galaxies picked by this definition are predominantly centrals, with $\sim10\%$ of isolated primaries being satellites \citep[see also][]{hwv15}. We consider isolated primaries in order to be close to the selection criteria used in recent observational studies \citep[see, e.g.,][]{kauffmann+13}. We investigate the impact of changing the selection criterion in Appendix~\ref{app:conftrends}.

The horizontal cyan line and associated band respectively show the mean and r.m.s. scatter over 10 realisations of the average all-galaxy red fraction which is independent of conformity strength. 
The mocks with $\rho = 0.01$ show nearly identical trends at $r\gtrsim2$Mpc regardless of the colour of the isolated primary, although the red fractions remain above the global value even as far as $10$Mpc from the center. There is a weak, conformity-like signal at small separations, which we have checked is entirely due to averaging over halo mass and disappears when using galaxies in fixed bins of halo mass. Together with the apparent `negative conformity' seen for this set of mocks in the left panel, this reiterates the need to be cautious when interpreting trends in analyses that average over luminosity and halo mass. 

Unlike the left panel, the two sets of mocks with $\rho = 0.99$ now show distinct trends. In the \emph{no-2h} mocks, the red fractions around blue and red isolated primaries are different until a distance of $\sim4$Mpc from the center (the one around red isolated primaries being higher), beyond which they are identical and also coincide with the ``no conformity'' red fractions associated with $\rho=0.01$. 
The default mocks, on the other hand, show a similar but substantially larger difference between the red fractions out to $\sim6$Mpc, beyond which they are also nearly identical but substantially larger than the red fractions in the other data sets (see section~\ref{sec:discuss} and Appendix~\ref{app:conftrends} for a discussion of these trends).

\begin{figure}
\centering
\includegraphics[width=0.45\textwidth]{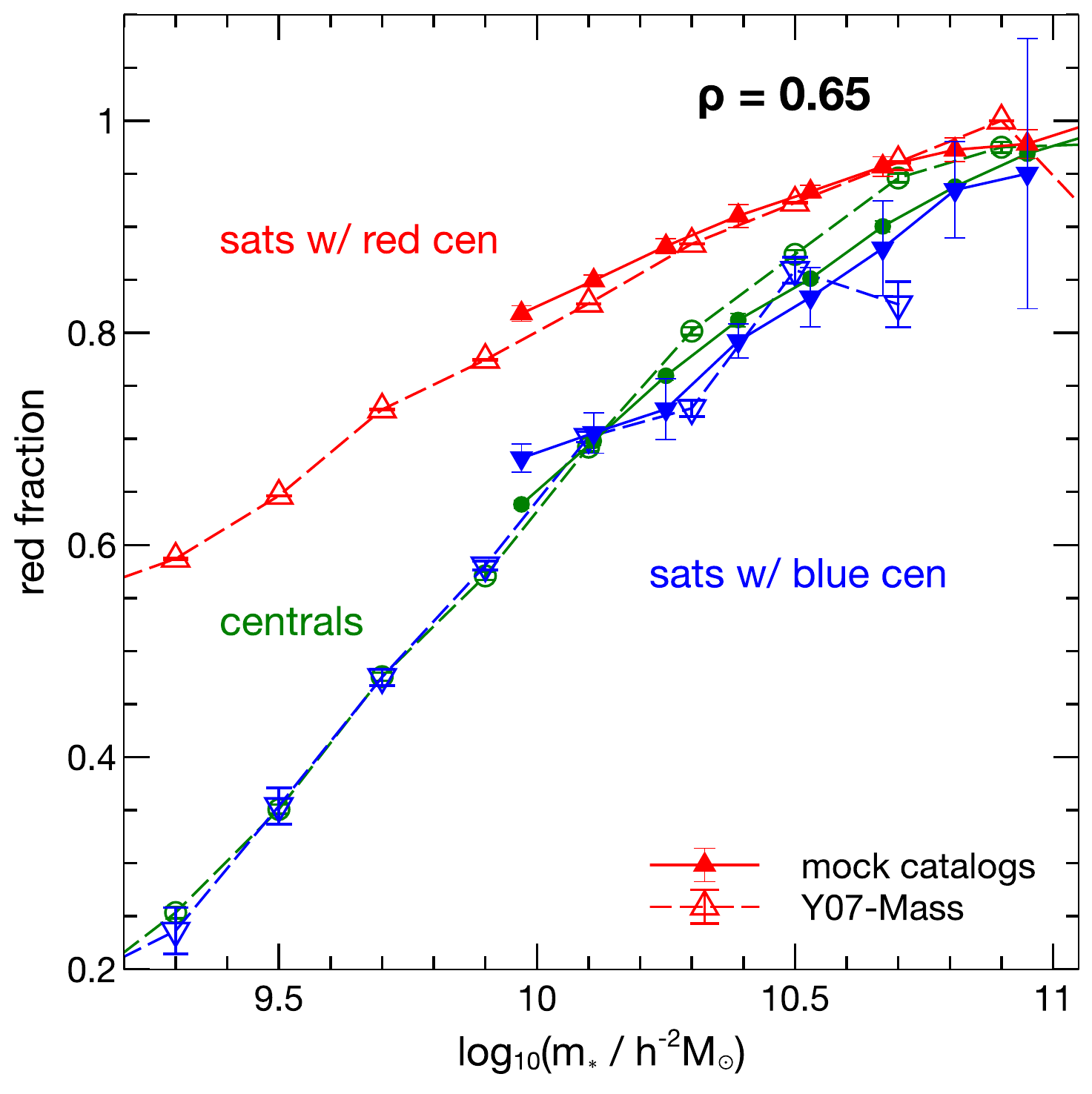}
\caption{Comparison of ``conformity red fractions'' $f_{\rm red|sat,rc}$ and $f_{\rm red|sat,bc}$ of satellites having red and blue centrals, respectively, with the red fraction of centrals $f_{\rm red|cen}$, for the Y07-Mass catalog (open symbols joined by dashed lines) and in mocks that used $\rho=0.65$ (filled symbols with solid lines). The red triangles, blue inverted triangles and green circles respectively show $f_{\rm red|sat,rc}$, $f_{\rm red|sat,bc}$ and $f_{\rm red|cen}$ (the last are the same as in Figure~\ref{fig:redfrac-avg}). The error bars on the mock results show the r.m.s. fluctuations around the mean value over $10$ realisations, while the error bars on the Y07 data are estimated from 150 bootstrap resamplings. Notice the remarkable similarity between $f_{\rm red|sat,bc}$ and $f_{\rm red|cen}$, especially at small masses in the Y07-Mass catalog.}
\label{fig:redfrac-conformity-1h}
\end{figure}

Figure~\ref{fig:redfrac-cc} shows the same quantities as Figure~\ref{fig:redfrac-2hconf}, except that the filled red crosses and blue stars joined by dashed lines now show the results for default mocks where we set $\rho=0.5$ (medium correlation). This plot shows how the difference between the red fractions around blue and red centrals/isolated primaries changes with group quenching efficiency $\rho$ and demonstrates the tunability of our model, which is one of our key results. 

\begin{figure}
\centering
\includegraphics[width=0.45\textwidth]{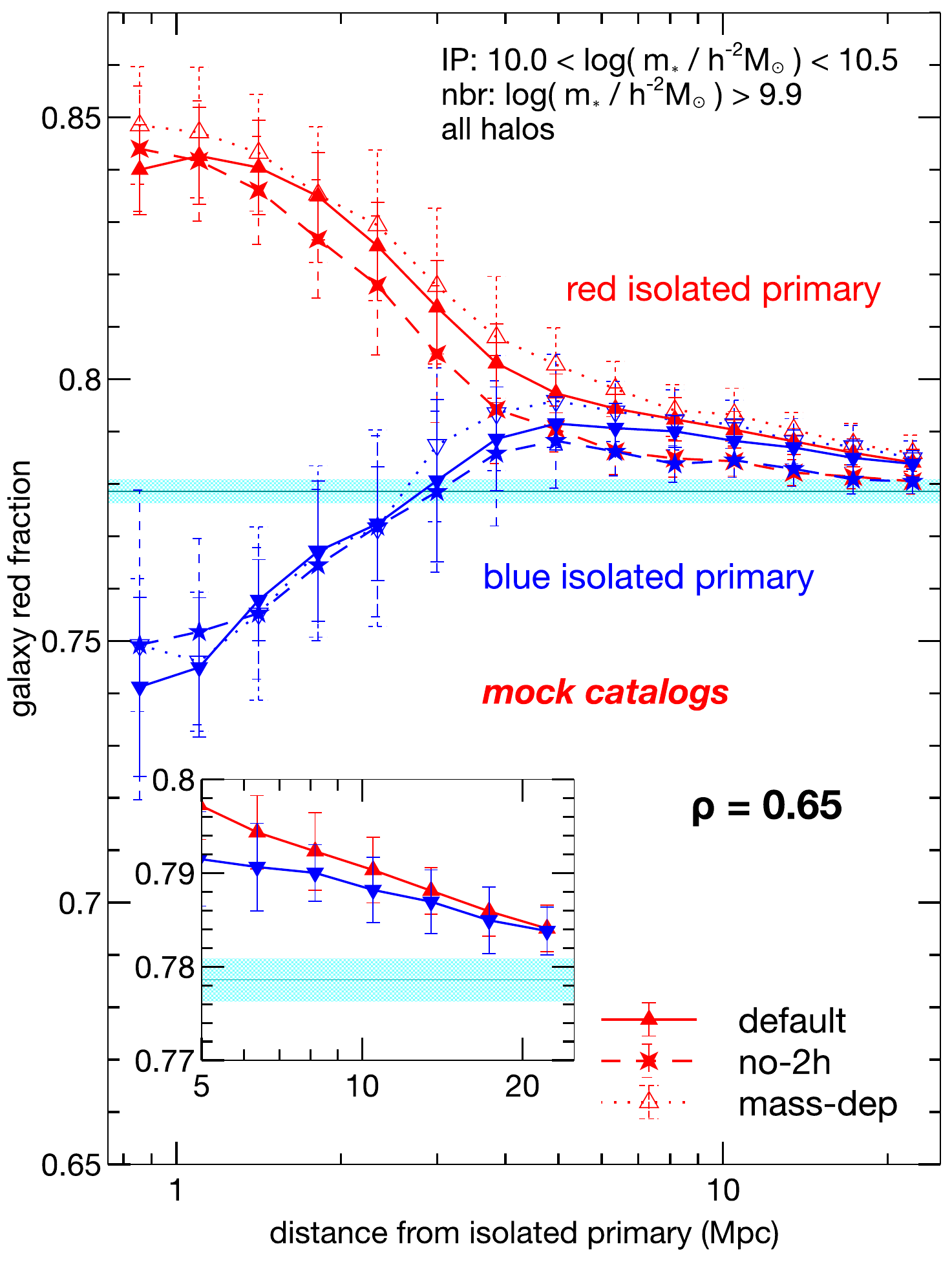}
\caption{Large scale conformity predicted in three configurations of mocks that used $\rho=0.65$ and whose respective 1-halo conformity signals are each consistent with measurements in the Y07 catalog. Filled symbols (red triangles and blue inverted triangles) with solid lines used our default mocks with halo mass-independent satellite and central red fractions at fixed luminosity, in which 2-halo conformity is switched on. Filled symbols (red crosses and blue stars) with dashed lines used the \emph{no-2h} mocks in which 2-halo conformity is switched off. Finally, open symbols (red triangles and blue inverted triangles) with dotted lines used mocks in which 2-halo conformity was switched on and the red fraction of centrals at fixed luminosity had an additional dependence on halo mass as described in Appendix~\ref{app:massdepredfrac}. The error bars on the mock results show the r.m.s. fluctuations around the mean value over $10$ realisations for each configuration. The inset zooms in on the behaviour of the signal in the default mocks at very large separations. Unlike Figures~\ref{fig:redfrac-2hconf} and~\ref{fig:redfrac-cc}, we restrict the isolated primaries to lie in the stellar mass range $10.0 < \log_{10}(m_\star) < 10.5$, while their neighbours are required to have $\log_{10}(m_\star) > 9.9$. The horizontal cyan line and band respectively show the average all-galaxy red fraction above the latter mass threshold and its r.m.s. scatter over 10 independent mocks (these values are independent of $\rho$).}
\label{fig:redfrac-conformity-2h}
\end{figure}

\subsection{Fixing the level of conformity}
\label{sec:results:subsec:fixconf}
\noindent
We notice in the left panel of Figure~\ref{fig:redfrac-cc} that the red fraction $f_{\rm red|sat,bc}$ of satellites with blue centrals is particularly sensitive to the value of $\rho$. This happens, at least in part, due to the inherent asymmetry of satellite colours, most of which are red.
Consequently, we can use measurements of 1-halo conformity in the Y07 catalog to fix the value of $\rho$, which can then be used to \emph{predict} the 2-halo conformity signal. 

By trial and error we have found that setting $\rho$ within $\sim10\%$ of $\rho=0.65$ leads to a behaviour of $f_{\rm red|sat,bc}$ in the mocks that closely resembles the one in the data over the mass range allowed by the mocks. 
Figure~\ref{fig:redfrac-conformity-1h} shows red fractions averaged over $10$ mocks that used $\rho=0.65$, and we see that both $f_{\rm red|sat,bc}$ and $f_{\rm red|sat,rc}$ agree very well with the Y07 measurements. The results of the previous section show that this agreement is independent of whether we use the default or the \emph{no-2h} model. Notice that the red fraction of satellites with blue centrals $f_{\rm red|sat,bc}$ in both data and mocks is remarkably similar to the corresponding $f_{\rm red|cen}$. This is particularly true for the data at low masses where our current mocks do not reach. 

Figure~\ref{fig:redfrac-conformity-2h} shows the large scale conformity signal predicted in the mocks having $\rho=0.65$, using both the default and the \emph{no-2h} models.
To assess the effect of our assumption of mass-independent satellite and central red fractions, we also show the corresponding signal in mocks having 2-halo conformity, but in which the red fraction of centrals is mass-\emph{dependent} as described in Appendix~\ref{app:massdepredfrac}. These were constructed to match the conformity-independent observables discussed earlier, as well as the 1-halo conformity signal, for which it suffices to use $\rho=0.65$ again (see Figure~\ref{fig:redfrac-conformity-mdep-1h}). 
In each case, we restrict the stellar mass of the isolated primaries to lie in the range $10.0 < \log_{10}(m_\star) < 10.5$.
We see that the overall signal in all three mocks is very similar at scales $\lesssim4$Mpc. At larger scales, the mocks with a halo mass dependent central red fraction show a slightly higher signal than in our default mocks. The large scale signal in the \emph{no-2h} mocks, on the other hand, is noticeably smaller than in the default. The inset zooms in on the behaviour of the signal in the default mocks at very large separations. The error bars depict the r.m.s. fluctuations around the mean over $10$ realisations for each configuration.

\begin{figure*}
\centering
\includegraphics[width=0.9\textwidth]{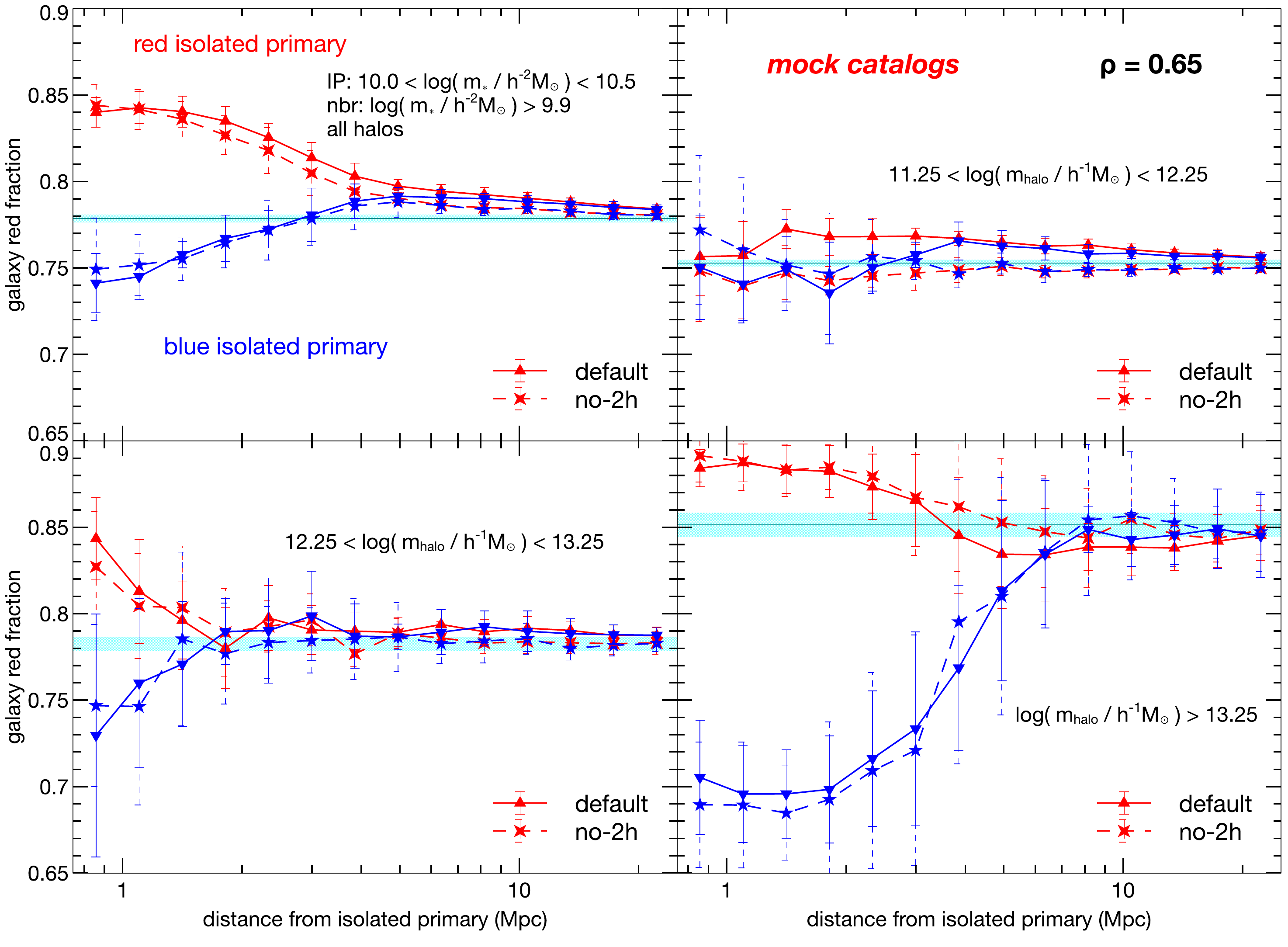}
\caption{Large scale conformity signal for $\rho=0.65$, broken down by halo mass as indicated. Note that the mass restrictions apply to the parent halos of both the isolated primary as well as its neighbours. Each panel is formatted identically to Figure~\ref{fig:redfrac-conformity-2h} and shows the results for the default and \emph{no-2h} mocks. See text for a discussion.}
\label{fig:redfrac-conformity-2h-mhbins}
\end{figure*}

Figure~\ref{fig:redfrac-conformity-2h-mhbins} shows a break-down of these large scale trends as a function of halo mass. Each panel shows the results for our default and \emph{no-2h} mocks (formatted identically to Figure~\ref{fig:redfrac-conformity-2h} and with the same restrictions on stellar masses of the isolated primaries and their neighbours), with the top left panel showing the same configuration as Figure~\ref{fig:redfrac-conformity-2h} and the top right, bottom left and bottom right panels showing bins of increasingly larger halo mass. Each panel shows the average all-galaxy red fraction in the respective halo mass bin, and its r.m.s. scatter, as the horizontal cyan line and band, respectively. We clearly see that the large difference between red fractions at separations $\lesssim4$Mpc arises primarily from large halo masses, while the distinction between the default and \emph{no-2h} signal at $\gtrsim8$Mpc only emerges at the smallest halo masses. 
We discuss these results in section~\ref{sec:discuss} below. 

\section{Discussion}
\label{sec:discuss}
\noindent
The 1-halo conformity trends seen in the left panels of Figures~\ref{fig:redfrac-2hconf} and~\ref{fig:redfrac-cc}, and in the mock results in Figure~\ref{fig:redfrac-conformity-1h} are a straightforward consequence of our choice of implementation and the definition of the group quenching efficiency $\rho$ as discussed in section~\ref{sec:algo:subsec:ccc}. As $\rho$ increases and approaches unity in our model, red (blue) satellites preferentially live in halos hosting red (blue) centrals. 
An interesting aspect of Figure~\ref{fig:redfrac-conformity-1h} is the remarkable similarity of the red fraction of satellites with blue centrals and the red fraction of centrals at low stellar mass. A similar result was obtained by \citet{phillips+14}, who found that the SFRs of satellites with star forming centrals are essentially the same as those of isolated galaxies with similar stellar mass. This would imply that the quenching efficiencies $\varepsilon_{\rm sat|SFc}$ of the satellites with star forming centrals are consistent with being zero.
On the other hand, \citet{klwk14} reported values of $\varepsilon_{\rm sat|SFc}$ different from zero. This apparent discrepancy is possibly due to differences in the sample definitions in these two studies; in particular, unlike \citet{phillips+14}, \citet{klwk14} limited their analysis to satellites residing in groups with at least $3$ members with $m_\star > 10^{10}\Msun$, thereby removing many low mass groups from the sample. As \citet{klwk14} also report a strong positive dependence of $\varepsilon_{\rm sat}$ on halo mass, it is natural to expect that the quenching efficiency in a sample biased toward more massive haloes would be nonzero. It will be very interesting to extend our mocks to lower stellar masses using a higher resolution simulation and compare the resulting 1-halo conformity trends -- particularly those at fixed group richness and halo mass -- with the data, but this is beyond the scope of the present work.

Turning to the large scale trends of red fractions surrounding isolated primaries, we encounter an even richer picture. It has been argued in the recent literature that large scale differences between red fractions or mean star formation rates surrounding blue and red isolated primaries must arise from environmental correlations across different halos \citep{kauffmann+13} and are therefore a signal of 2-halo conformity \citep{hwv15}. Our results in Figure~\ref{fig:redfrac-2hconf} and especially Figure~\ref{fig:redfrac-conformity-2h}, however, show that an alternative explanation is also possible.
Figure~\ref{fig:redfrac-conformity-2h} used three different sets of 
mocks, \emph{all} of which are constructed to give a 1-halo conformity signal consistent with measurements in the Y07 catalog. Only two of these mocks have a genuine 2-halo conformity built in, however; the third contains randomly assigned halo concentrations which are not correlated with the halo environment. This third set of \emph{no-2h} mocks serves as a useful toy example of the possibility that conformity might arise due to some property of halos that \emph{does not} exhibit large scale environmental effects (see the discussion in the Introduction). The large scale conformity signal for all three mocks is \emph{very similar at scales} $\lesssim4$Mpc.
Figure~\ref{fig:redfrac-conformity-2h-mhbins} shows that the difference of red fractions at these scales is dominated by halo masses $\log_{10}(m)\gtrsim13.25$ (bottom right panel) -- for which it is consistent with being a 1-halo effect\footnote{For comparison, a cluster-sized halo of mass $10^{14}\Mh$ has a radius $R_{\rm 200b}\simeq1.7$Mpc for our cosmology, and the largest halos in our simulations reach masses of $\sim10^{14.8}$-$10^{15.2}\Mh$.} -- and has a smaller and noisier contribution from small halo masses (top right panel, see also Appendix~\ref{app:conftrends}).
This means that, at these scales, we could easily mistake a purely 1-halo effect for genuine 2-halo conformity\footnote{Figure~\ref{fig:redfrac-conformity-2h} also shows that the large scale conformity signal appears to be robust against a mass dependence of the central red fraction, showing a slight increase on average as compared to the default mocks.}
Similar conclusions can be drawn by comparing the $\rho=0.99$ measurements in the right panel of Figure~\ref{fig:redfrac-2hconf} with those in the mocks with $\rho=0.01$ or no conformity. The $\rho=0.99$ \emph{no-2h} measurements are identical to the $\rho=0.01$ measurements at large scales, which is expected since neither of these mocks has any 2-halo conformity. The measurements in the two $\rho=0.99$ mocks on the other hand are very similar at scales $\lesssim4$Mpc. Taken together, this shows that that the conformity signal at $\lesssim4$Mpc in the $\rho=0.99$ mocks is almost entirely a 1-halo effect (see Appendix~\ref{app:conftrends} for further discussion).

From the observational point of view, halo mass-dependencies can in principle be controlled by performing tests at, e.g., fixed group richness \citep{koester+07a,ssm07,skibba09}, stellar mass of the central \citep{more+11} or any other observable property that correlates well with halo mass \citep{lys15}. 
Notice, however, that all the measurements in Figures~\ref{fig:redfrac-conformity-2h} and~\ref{fig:redfrac-conformity-2h-mhbins} use isolated primaries with stellar masses $10.0 < \log_{10}(m_\star) < 10.5$ and still cannot distinguish between the default and \emph{no-2h} models at scales $\lesssim4$Mpc, unless the halo mass is explicitly restricted to small values. We have also checked that the relative behaviour of our models at these scales remains unchanged when using primaries of different stellar mass\footnote{ Figure~\ref{fig:redfrac-conformity-2h-smbins} in the Appendix shows that the signal, particularly at large scales, is essentially independent of the stellar mass of the primary; we attribute this to a large scatter in the distribution of halo masses at fixed stellar mass (Figure~\ref{fig:SHMR}).}.
This is unfortunate, since it very likely means that one cannot conclusively argue that the conformity measured by \citet{kauffmann+13} at projected scales $\lesssim4$Mpc in similar mass bins is evidence of galaxy assembly bias. Projection effects are unlikely to alter this conclusion. It will be useful to repeat such tests at fixed group richness, which might be a better indicator of halo mass.

For now, it is very interesting to note that there \emph{is}, in fact, a signal which diminishes in the absence of genuine 2-halo conformity and is also robust against the inclusion of halo mass dependence in the red fractions. This is the fact that, in the presence of 2-halo conformity, the red fractions at \emph{even larger} scales ($\gtrsim8$Mpc) surrounding both red and blue isolated objects \emph{remain above the global average value} by a small but statistically significant amount, showing nearly identical values and a trend that is relatively insensitive to the halo mass dependence of the central red fraction. When genuine 2-halo conformity is \emph{absent}, the red fractions at these scales are discernably closer to the global average; as noted earlier, they are also identical to the ``no conformity'' red fractions at these scales. These trends can be seen in Figures~\ref{fig:redfrac-2hconf},~\ref{fig:redfrac-conformity-2h} and~\ref{fig:redfrac-trugrps}, with the elevation compared to the global value seen clearly in the zoom-in inset panel of Figure~\ref{fig:redfrac-conformity-2h}. 
Figure~\ref{fig:redfrac-conformity-2h-mhbins} shows that this signal is absent at large halo masses and only emerges in the range $11.25 < \log_{10}(m) < 12.25$.

This dependence on halo mass is exactly what one expects from a signal driven by halo assembly bias; in fact, one can also derive a more detailed understanding of the \emph{nature} of the signal using the Halo Model. In Appendix~\ref{app:conftrends}, e.g., we argue that the red fractions at very large separations from central galaxies are determined by ratios of cross-correlation functions of red/blue galaxies with red/blue centrals, in which the concentration-dependence of halo assembly bias couples with that of galaxy colour introduced in our default model and leads to an elevation that is qualitatively similar to what we see in Figure~\ref{fig:redfrac-conformity-2h}.
The fact that the signal is only of order a few percent is consistent with halo assembly bias being a weak effect. 

Our model assumes that central galaxies are always the brightest members of their respective groups, and we have also imposed this condition when defining centrals in the Y07 catalog. A number of studies indicate that this is not a good assumption for a substantial fraction of groups \citep[see, e.g.,][]{skibba+11,masaki+13,hmts13}. Since galactic conformity manifests as a \emph{similarity} of galaxy colours in groups, any errors in classifying galaxies as centrals and satellites, provided they have a minimal impact on overall group membership, will tend to make these populations similar to each other and are therefore likely to \emph{induce} conformity-like effects \citep{campbell+15}. Consequently, it is possible that our analysis in Section~\ref{sec:results:subsec:fixconf} somewhat overestimates the strength of conformity (i.e., the value of $\rho$). However, the results of Section~\ref{sec:results:subsec:conf} suggest that our conclusions regarding the \emph{relative} strengths of the conformity signal at various scales are qualitatively robust to such errors. It will nevertheless be interesting to revisit this issue in future work using more refined criteria for identifying centrals.

A proper comparison with data will require accounting for projection effects both in the red fraction measurements as well as the definition of isolated primaries themselves. Although these will dilute the signal, its strength should also increase upon including galaxies with smaller stellar masses (and stronger assembly bias) than we could access in our mocks, and overall we expect that the difference between large scale 1-halo and genuine 2-halo effects will remain measurable. Additionally, we expect that simultaneous measurements of 1-halo and large scale conformity together with colour-dependent clustering will be required to break the weak degeneracy between halo mass dependence of the central red fraction and the level of conformity. Such a joint analysis will also be important from the point of view of obtaining unbiased HOD fits that account for galactic conformity \citep{zhv14}. 


\section{Summary and Conclusions}
\label{sec:conclude}
\noindent
We have introduced a flexible model of galactic conformity within the Halo Occupation Distribution (HOD) framework by modifying and extending the algorithm described by \citet[][]{ss09}. By construction, our mock galaxy catalogs show good agreement with measurements of conformity-\emph{independent} variables in the \citet[][Y07]{yang+07} group catalog based on DR7 of the SDSS. These variables include the satellite fraction, the red fractions of all galaxies, centrals and satellites, and the satellite quenching efficiency $\varepsilon_{\rm sat}$ (Figure~\ref{fig:redfrac-avg}), as well as the all-galaxy luminosity and stellar mass functions (Figure~\ref{fig:massfuncs}). Galaxy luminosities are assigned using the HOD \citep[we use the calibration by][]{zehavi+11}, colours are assigned using colour-luminosity fits to SDSS data (Figure~\ref{fig:gr-comparemock} and equations~\ref{dbl-Gauss-fits}) and stellar masses are assigned using a colour-dependent mass-to-light ratio (equation~\ref{masstolight-fit}), also fit to SDSS data. Our mock catalogs are luminosity-complete for $M_r<-19.0$ and mass-complete for $\log_{10}(m_\star)\gtrsim9.9$ (Figures~\ref{fig:massfuncs} and~\ref{fig:hist2d-cen}).

Our algorithm introduces conformity between the colours of the central and satellites of a group (a 1-halo effect) by using a tunable group quenching efficiency $\rho$ to correlate these colours with the concentration of the parent dark halo of the group. Halo concentration (which has a scatter at fixed halo mass) is therefore identified as the ``hidden variable'' in our model which causes galactic conformity even in halos of fixed mass. Halo assembly bias then leads to a 2-halo effect at very large scales (this is our default model), which we can also switch off by randomizing halo concentrations among halos of fixed mass (we call this the \emph{no-2h} model). The latter is a useful toy example in which conformity arises due to some unspecified property of halos (e.g., this might be a coupling between star formation activity and the hot gas content in a halo) that \emph{does not} exhibit large scale environmental effects.
We have performed various tests to study the nature of the signal, including changing $\rho$ (Figure~\ref{fig:redfrac-cc}) in the presence or absence of 2-halo conformity (Figure~\ref{fig:redfrac-2hconf}) for different choices of isolation criteria -- we used isolated primaries (Figures~\ref{fig:redfrac-2hconf} and~\ref{fig:redfrac-cc}) as well as group centrals (Figure~\ref{fig:redfrac-trugrps}).
Our main results can be summarized as follows.
\begin{itemize}
\item We find that setting $\rho=0.65$ gives a 1-halo conformity signal in the mocks which agrees well with the corresponding signal in the Y07 catalog (Figure~\ref{fig:redfrac-conformity-1h}). The signal manifests as a difference between the red fractions of satellites in groups with red and blue centrals. 
Additionally, in the Y07 catalog, the red fraction of satellites with blue centrals is remarkably similar to the red fraction of centrals at all masses; our mocks correctly reproduce this trend down to their completeness limit.
\item The above value of $\rho$ also leads to a signal at large scales in the mocks (Figure~\ref{fig:redfrac-conformity-2h}); specifically, we see a significant difference between the galaxy red fractions surrounding red and blue isolated primaries out to separations $\lesssim4$Mpc. Interestingly, we find that this signal is dominated by the 1-halo contribution of large halos with $\log_{10}(m) > 13.25$ (Figure~\ref{fig:redfrac-conformity-2h-mhbins}), and \emph{persists even when we switch off 2-halo conformity} in our \emph{no-2h} model. This implies that the observation of such a difference \citep[e.g.][]{kauffmann+13} is not conclusive evidence that galactic conformity arises from halo assembly bias, since it could also arise from 1-halo effects ``leaking'' to large scales due to averaging over halo mass (section~\ref{sec:discuss}). 
\item At even larger scales ($\gtrsim8$Mpc in Figure~\ref{fig:redfrac-conformity-2h}), the signals with and without 2-halo conformity \emph{do} become distinct, with the genuine 2-halo signal remaining \emph{significantly elevated} compared to the global average red fraction out to separations in excess of $15$Mpc (at least in the 3-d case that we consider). 
This 2-halo signal is absent at large halo masses and only emerges at smaller masses $11.25 < \log_{10}(m) < 12.25$ (Figure~\ref{fig:redfrac-conformity-2h-mhbins}), being qualitatively consistent with expectations from halo assembly bias (section~\ref{sec:discuss} and Appendix~\ref{app:conftrends}).
We therefore suggest that this elevation compared to the global average could be a more robust indicator of large scale galaxy assembly bias than is the difference between red fractions at scales $\lesssim4$Mpc.
\end{itemize}

We end with a brief discussion of future extensions of our work.
In a forthcoming paper (Kova\v c et al., in preparation), we will compare the large scale signal in our mocks, after accounting for projection effects, with a corresponding measurement in the Y07 catalog to determine whether or not the observed large scale conformity is due to galaxy assembly bias or a residual of some other 1-halo process. Tests at fixed group richness will also be useful in answering this question. 
It will also be interesting to explore the use of galaxy lensing and/or traditional correlation function analyses to validate the connection between galaxy colours and host halo concentrations assumed in this work, particularly to check for consistency with the value of $\rho$ presented above.
In another publication (Pahwa et al., in preparation), we will present the analytical formalism for including conformity in the HOD framework; we will show that this requires straightforward modifications in existing HOD pipelines. 

Our algorithm can also be extended to include radial profiles for satellite velocity dispersions and colours \citep{prescott+11,hartley+15}, as well as concentration-dependent satellite abundances \citep{wechsler+06,mww15}. The signal at high redshift is interesting too; this could be modelled using analytical prescriptions tuned to match high-redshift luminosity function and clustering data \citep{ttc13,bwc13,jss14}.
Finally, it will be extremely interesting to extend our algorithm to lower stellar masses using accurate faint-end ($M_{r,{\rm max}}\sim-16.0$) HOD fits and higher resolution $N$-body simulations. We estimate that a simulation with $1024^3$ particles in a $(100\Mpch)^3$ box will allow us to create catalogs that are mass-complete for $\log_{10}(m_\star)\gtrsim8.7$; comparing these to data will provide us with stringent tests on the nature of the conformity signal.


\section*{Acknowledgements}
We thank J. Woo and M. Shirazi for help with the Kcorrect tools, and an anonymous referee for an insightful report that has helped improve the presentation.
We are grateful to the SDSS collaboration for publicly releasing their data set, and Yang et al. for releasing their updated group catalog.
We thank O. Hahn, V. Springel and P. Behroozi for making their codes publicly available. 
We gratefully acknowledge computing facilities at ETH, Z\"urich and IUCAA, Pune.
Our mock catalogs are available upon request.



\setlength{\bibhang}{2.0em}
\setlength\labelwidth{0.0em}
\bibliography{masterRef,mockRefs}


\appendix
\section{}
\subsection{The (in)dependence of galaxy red fractions on halo mass}
\label{app:massdepredfrac}
\noindent
As mentioned in the main text, one might expect in general that the distribution of galaxy colours depends on both galaxy luminosity and host halo mass. We explore this idea further in this Appendix.

Let the all-galaxy red fraction be some function $p({\rm red}|M_r,m)$ at fixed galaxy luminosity and parent halo mass. Separating the red fractions of centrals and satellites, we can write this as
\begin{align}
p({\rm red}|M_r,m)&=p({\rm red}|{\rm cen},M_r,m)\,p({\rm cen}|M_r,m) \notag\\
&\ph{p({\rm red})}
+ p({\rm red}|{\rm sat},M_r,m)\,p({\rm sat}|M_r,m)\,,
\label{cen-sat-split}
\end{align}
where $p({\rm sat}|M_r,m)$ gives the fraction of galaxies that are satellites [so that $p({\rm cen}|M_r,m)=1-p({\rm sat}|M_r,m)$], and $p({\rm red}|{\rm sat/cen},M_r,m)$ gives the fraction of satellites/centrals that are red. Any mass dependence in $p({\rm red}|{\rm sat/cen},M_r,m)$ could lead to ``conformity-like'' effects in analyses that average over halo mass (as we saw in Figure~\ref{fig:redfrac-2hconf}). What is usually referred to as conformity, however, is the statement that central and satellite colours are correlated even at \emph{fixed} halo mass. Since we are trying to model the latter effect, it is important to understand the former as well.

For ease of notation, we define the following quantities that are fixed by the HOD:
\begin{align}
\Cal{N}_{\rm sat}(<M_r|m) &\equiv f_{\rm cen}(<M_r|m)\bar N_{\rm sat}(<M_r|m)\notag\\
\Cal{N}_{\rm sat}(M_r|m) &\equiv \partial \Cal{N}_{\rm sat}(<M_r|m) / \partial M_r\notag\\
f_{\rm cen}(M_r|m) &\equiv \partial f_{\rm cen}(<M_r|m) / \partial M_r\,,
\label{HODderivs}
\end{align}
where $f_{\rm cen}(<M_r|m)$ and $\bar N_{\rm sat}(<M_r|m)$ were introduced in section~\ref{sec:algo:subsec:S09}.
The satellite fraction is then also fixed by the HOD as:
\be
p({\rm sat}|M_r,m) = \frac{\Cal{N}_{\rm sat}(M_r|m)}{f_{\rm cen}(M_r|m)+\Cal{N}_{\rm sat}(M_r|m)}\,,
\label{psat-massdep}
\ee
In principle, however, we must calibrate the luminosity and mass-dependence of both $p({\rm red}|{\rm sat},M_r,m)$ and $p({\rm red}|{\rm cen},M_r,m)$. The S09 algorithm simplifies the problem by postulating that these two functions are approximately independent of parent halo mass. In other words, the S09 model states
\begin{align}
p({\rm red}|M_r,m) &= p({\rm red}|{\rm cen},M_r)\,p({\rm cen}|M_r,m)\notag\\ 
&\ph{p({\rm red})}
+ p({\rm red}|{\rm sat},M_r)\,p({\rm sat}|M_r,m)\,,
\label{cen-sat-split-S09}
\end{align}
so that the all-galaxy red fraction still inherits a mass dependence from the HOD.
S09 proceed to calibrate the function $p({\rm red}|{\rm sat},M_r)$ using clustering measurements of SDSS galaxies, and fix $p({\rm red}|{\rm cen},M_r)$ by demanding that the halo-averaged version of \eqn{cen-sat-split-S09} return the measured all-galaxy red fraction as a function of luminosity:
\begin{align}
p({\rm red}|M_r) &= p({\rm red}|{\rm cen},M_r)\,\bar p({\rm cen}|M_r) \notag\\
&\ph{p({\rm red})}
+ p({\rm red}|{\rm sat},M_r)\,\bar p({\rm sat}|M_r)\,,
\label{cen-sat-split-avg}
\end{align}
where
\begin{align}
&\bar p({\rm sat}|M_r) = \frac{\int\der m\,n(m)\,\Cal{N}_{\rm sat}(M_r|m)}{\int\der m\,n(m)\,\left[f_{\rm cen}(M_r|m)+\Cal{N}_{\rm sat}(M_r|m)\right]}\,,
\label{pbarsat}
\end{align}
and $\bar p({\rm cen}|M_r) = 1-\bar p({\rm sat}|M_r)$, with $n(m)\der m$ being the number density of $m$-halos. 

\begin{figure}
\centering
\includegraphics[width=0.45\textwidth]{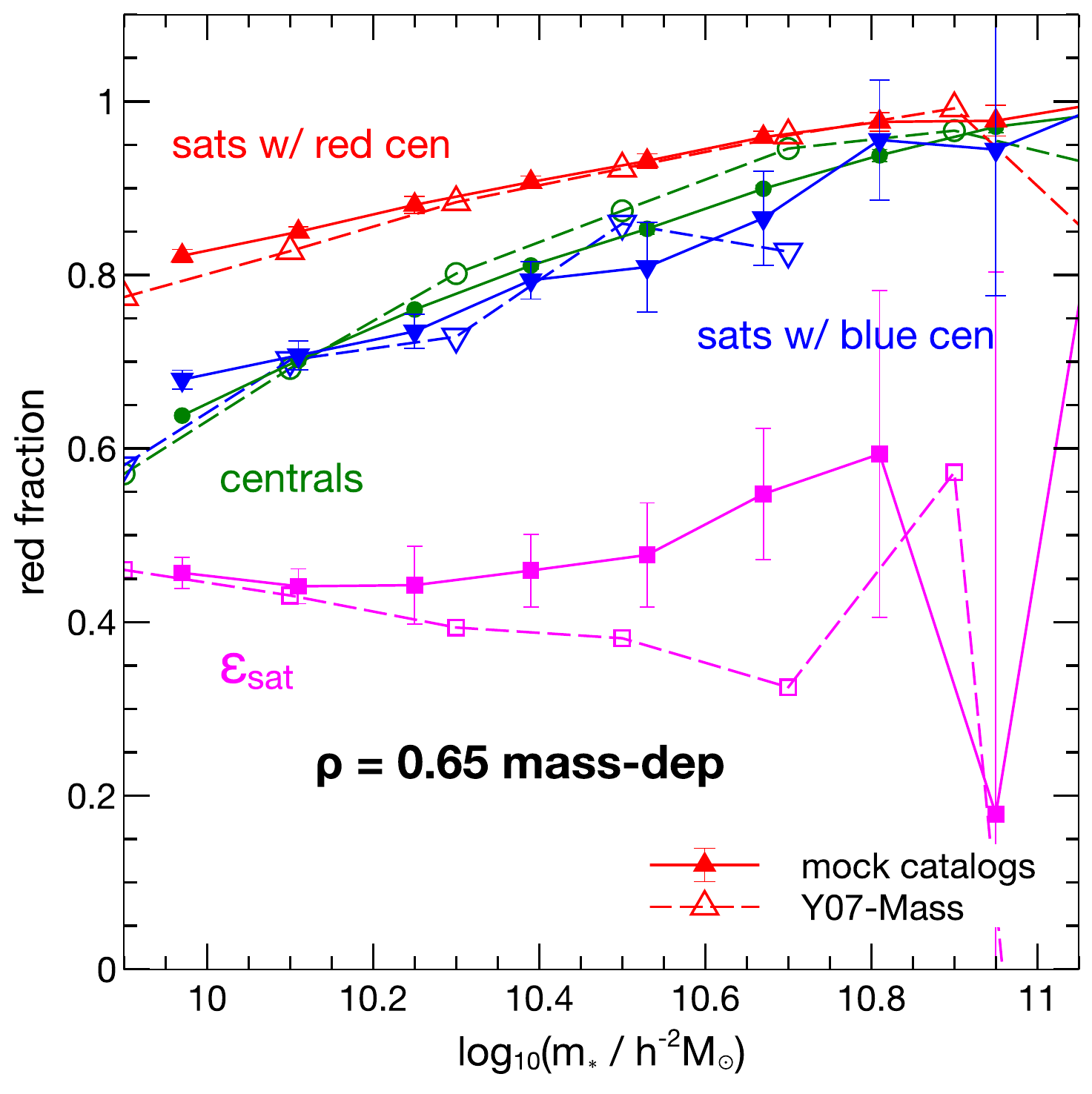}
\caption{Comparison of the conformity-independent observables $f_{\rm red|cen}$ (green circles) and $\varepsilon_{\rm sat}$ (magenta squares), and the conformity-dependent $f_{\rm red|sat,rc}$ (red triangles) and $f_{\rm red|sat,bc}$ (blue inverted triangles) in the Y07 catalog (open symbols with dashed lines) and in mocks that used $\rho=0.65$ and a halo mass dependent central red fraction given by \eqns{predcen-massdep}-\eqref{avg_massdep_factor} (filled symbols with solid lines). The error bars on the mock results show the r.m.s. fluctuations around the mean value over $10$ realisations.}
\label{fig:redfrac-conformity-mdep-1h}
\end{figure}

The assumptions underlying \eqn{cen-sat-split-S09} were tested by \cite{skibba09} by comparing colour distributions at fixed group richness as predicted by the S09 model with those in the Y07 and \citet{berlind+06} catalogs. Overall the data appear consistent with the assumption that the red fractions at fixed $M_r$ -- of both satellites \emph{and} centrals -- do not depend on halo mass. As S09 point out, this may be expected for satellites, whose \emph{luminosity} function is observed to be approximately halo mass-independent \citep{ssm07,hansen+09}; the situation for the centrals, however, is less clear since their luminosity function does depend strongly on halo mass, at least for small masses. 
Also, the calibration by S09 using clustering measurements ignored effects of conformity and is therefore difficult to interpret in the present context. It is conceivable that a weak halo mass dependence $p({\rm red}|{\rm cen},M_r,m)$ might have litte consequence for the fixed-richness tests of \cite{skibba09} but might have noticeable effects, e.g., in the red fraction profiles of Figures~\ref{fig:redfrac-2hconf} and~\ref{fig:redfrac-cc}. 

\begin{figure*}
\centering
\includegraphics[width=0.85\textwidth]{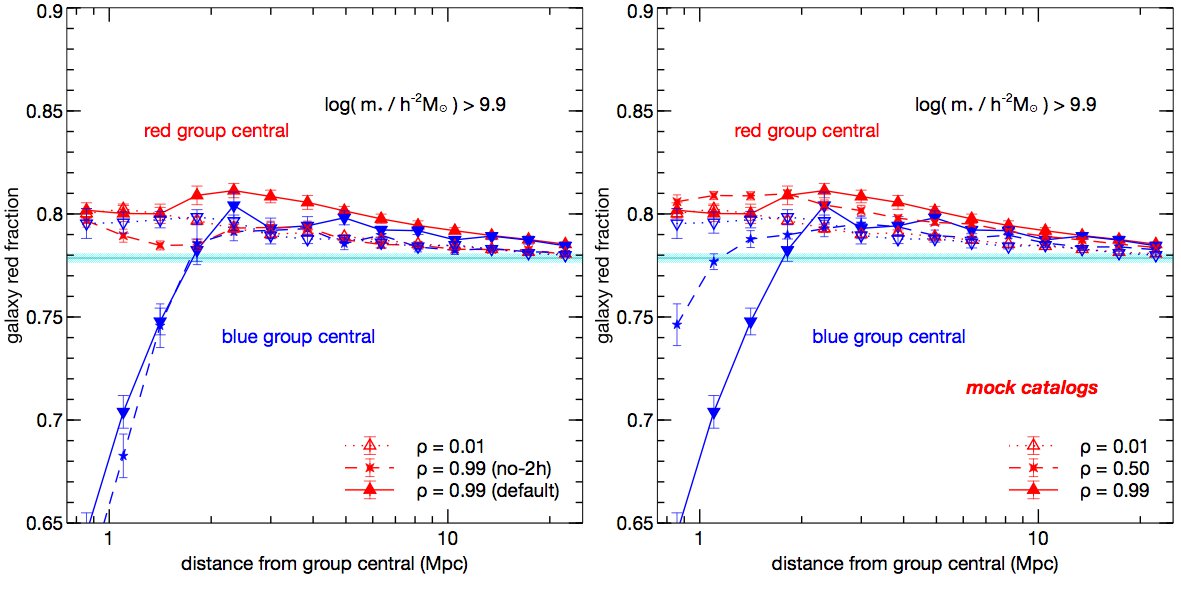}
\caption{Same as the right panels of Figures~\ref{fig:redfrac-2hconf} and~\ref{fig:redfrac-cc} using the same set of mocks, except that the galaxy red fractions are measured centered on group centrals (centrals with at least one associated satellite) rather than isolated primaries. See text for a discussion.}
\label{fig:redfrac-trugrps}
\end{figure*}

To assess the importance of such effects, we explore an alternative toy model which also assumes a mass-independent satellite red fraction, but has a mass-\emph{dependent} central red fraction $p({\rm red}|{\rm cen},M_r,m)$ whose average satisfies \eqn{cen-sat-split-avg}, with $p({\rm red}|{\rm sat},M_r)$ and $p({\rm red}|M_r)$ given by \eqn{p-redsat} and the first of \eqns{dbl-Gauss-fits}, respectively. At fixed luminosity, this model preferentially places red centrals in massive halos. In particular, denoting $\bar p_{\rm rc}\equiv p({\rm red}|{\rm cen},M_r)$, we use
\begin{align}
p({\rm red}|{\rm cen},M_r,m) &= \bar p_{\rm rc}\,\frac{g(M_r,m)}{\avg{g(M_r,m)}}\,,
\label{predcen-massdep}
\end{align}
where we set
\be
g(M_r,m) = \bar p_{\rm rc} + \left(1-\bar p_{\rm rc}\right)\,\frac{\tanh(m/M_{\rm min})}{\tanh(1)}\,,
\label{massdep_factor}
\ee
with
\be
\avg{g(M_r,m)} \equiv \frac{\int\der m\,n(m)\,f_{\rm cen}(M_r|m)\,g(M_r,m)}{\int\der m\,n(m)\,f_{\rm cen}(M_r|m)}\,,
\label{avg_massdep_factor}
\ee
and where $M_{\rm min}(M_r)$ is the mass scale where $f_{\rm cen}(<M_r|m)=1/2$. The model is chosen such that in the limit of a step-like function $f_{\rm cen}(<M_r|m) = \Theta(m-M_{\rm min}(M_r))$, we would have $\avg{g(M_r,m)} = 1$ and $p({\rm red}|{\rm cen},M_r,m=M_{\rm min}) = p({\rm red}|{\rm cen},M_r)$. The HOD we use has a smoother mass dependence in $f_{\rm cen}(<M_r|m)$ and therefore our model requires a renormalisation by $\avg{g(M_r,m)}$. The model ensures $0 < p({\rm red}|{\rm cen},M_r,m) < 1$ in practice for all $M_r$ and $m$. 

Figure~\ref{fig:redfrac-conformity-mdep-1h} shows the conformity-dependent satellite red fractions as well as the central red fraction and $\varepsilon_{\rm sat}$ when setting $\rho=0.65$ in this model as the filled symbols joined by solid lines, formatted as in Figures~\ref{fig:redfrac-avg} and~\ref{fig:redfrac-conformity-1h}. The open symbols with dashed lines again show corresponding measurements in the Y07-Mass catalog, and we see a good agreement between the mocks and the data. The 2-halo signal predicted by this model is shown in Figure~\ref{fig:redfrac-conformity-2h} as the open symbols/dashed lines and is discussed in section~\ref{sec:discuss}.


\subsection{Large scale trends with and without conformity}
\label{app:conftrends}
\noindent
Here we further explore the nature of large scale conformity in our mocks and argue that the trends seen in Figures~\ref{fig:redfrac-2hconf} and~\ref{fig:redfrac-cc} can be understood as a consequence of the relative bias of blue and red galaxies, together with the nature of halo assembly bias.

It is interesting to compare the results in the right panels of Figures~\ref{fig:redfrac-2hconf} and~\ref{fig:redfrac-cc} with similar measurements performed around ``group centrals'' rather than isolated primaries, where a group central is a central with at least one associated satellite. Firstly, we find that the number density of group centrals in our simulation box ($\sim1.1\times10^{-3} (h/{\rm Mpc})^3$) is only about $16\%$ that of isolated primaries, meaning that isolated primaries predominantly comprise of isolated \emph{singletons} which occupy low mass halos, which is not surprising considering that $\sim90\%$ of all centrals are singletons. 
The all-galaxy red fraction surrounding group centrals is shown in Figure~\ref{fig:redfrac-trugrps}, whose left and right panels are formatted exactly like the right panels of Figures~\ref{fig:redfrac-2hconf} and~\ref{fig:redfrac-cc}, respectively. 
We see, e.g., that the magnitude of the difference between the $\rho=0.99$ red fractions around red and blue group centrals is substantially reduced as compared to that around isolated primaries. 
At scales larger than $\sim15$Mpc, the \emph{no-2h} mocks show red fractions consistent with the global average, while the default mocks with increasingly larger $\rho$ show correspondingly larger red fractions.

To understand these trends, consider the simpler measurement of the galaxy red fraction at a distance $r$ from an \emph{arbitrary} red central, $f_{\rm red|rc}(r)$. A straightforward calculation shows that we can write this as
\be
f_{\rm red|rc}(r) = \left(1 + \left(\frac{\bar n_{\rm b}(1+\xi_{\rm b,rc}(r))}{\bar n_{\rm r}(1+\xi_{\rm r,rc}(r))}\right)\right)^{-1}\,,
\label{analytic-fred|rc(r)}
\ee
where $\bar n_{\rm b(r)}$ is the mean number density of blue (red) galaxies and $\xi_{\rm b(r),rc}(r)$ is the cross-correlation function between blue (red) galaxies and red centrals. A similar expression holds for the red fraction around blue centrals $f_{\rm red|bc}(r)$, with the replacement `rc'$\to$`bc' in the correlation functions on the r.h.s. of \eqn{analytic-fred|rc(r)}. 

The correlation functions above can be split into contributions arising from galaxies in the same halo as the central object or in different halos (the 1-halo and 2-halo terms, respectively): $\xi(r) = \xi^{\rm (1h)}(r) + \xi^{\rm (2h)}(r)$. Recall that 1-halo terms dominate at small separations and 2-halo terms at large separations. For sufficiently large separations the 2-halo terms also become small compared to unity.
We now observe the following:

\begin{figure*}
\centering
\includegraphics[width=0.45\textwidth]{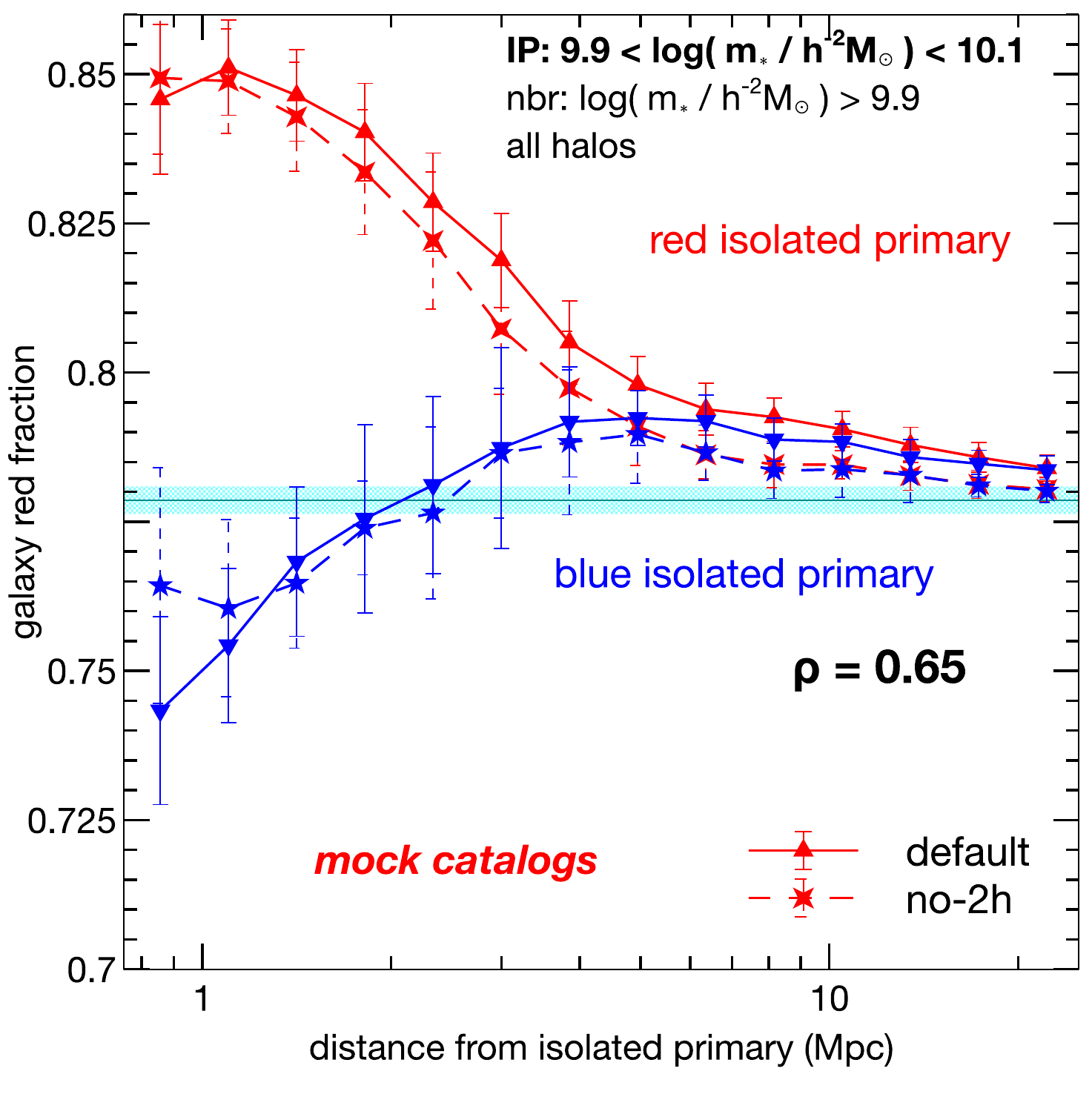}
\includegraphics[width=0.45\textwidth]{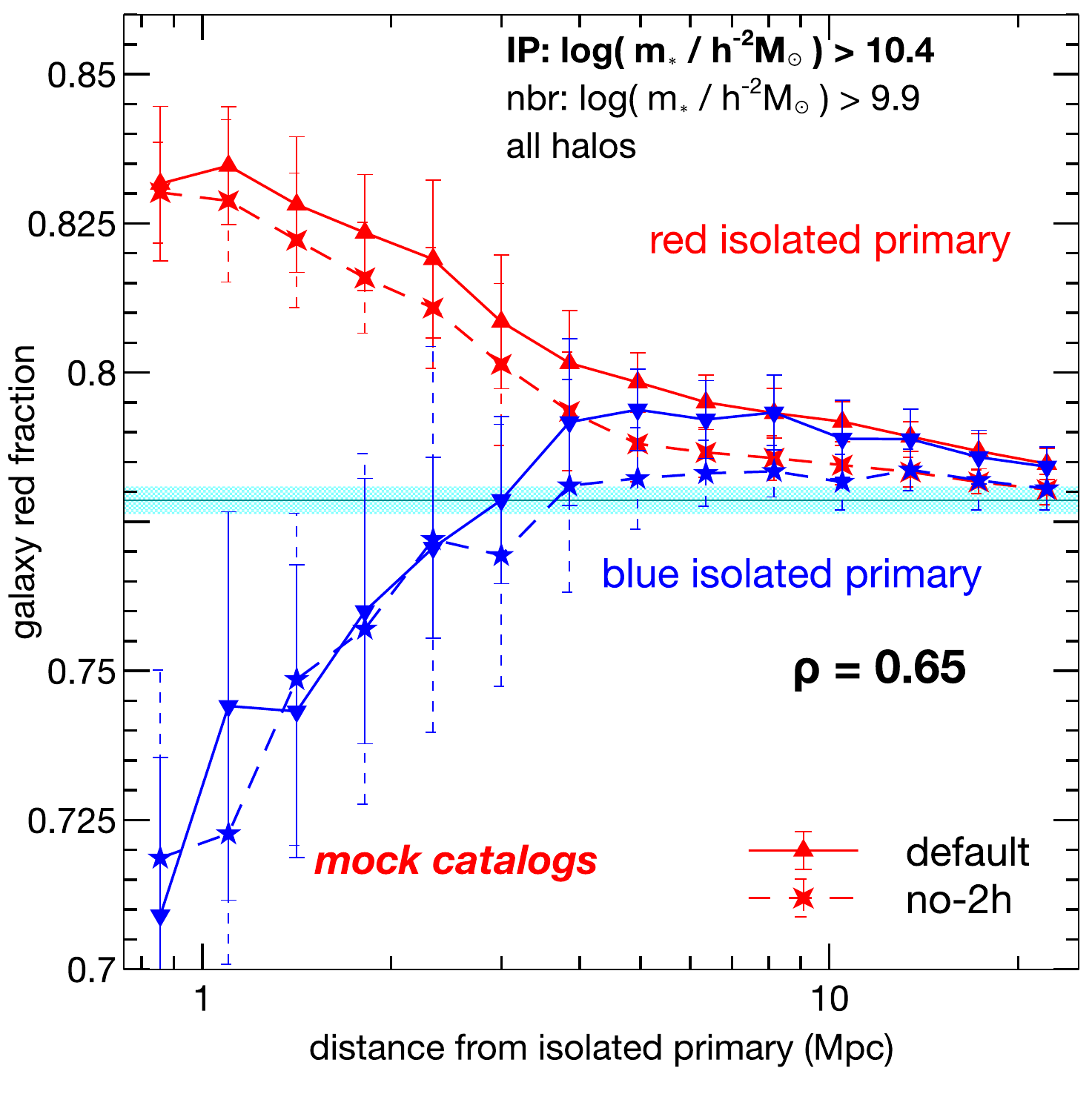}
\caption{Same as Figure~\ref{fig:redfrac-conformity-2h}, showing the large scale conformity signal for the default and \emph{no-2h} mocks around isolated primaries with $9.9<\log_{10}(m_\star)<10.1$ \emph{(left panel)} and $\log_{10}(m_\star)>10.4$ \emph{(right panel)}.}
\label{fig:redfrac-conformity-2h-smbins}
\end{figure*}

\begin{itemize}
\item {\bf Red fraction profile:}\\ 
At scales large enough that the 2-halo contributions dominate but are not yet negligible compared to unity, $f_{\rm red|rc}(r)$ will have an $r$-dependence given by replacing $\xi(r)\to\xi^{\rm (2h)}(r)$ in \eqn{analytic-fred|rc(r)}. In general, the correlation functions would not conspire to precisely cancel the $r$-dependence in this regime. Therefore, only when \emph{both} 2-halo correlations have become substantially smaller than unity will the asymptotic value $f_{\rm red|rc}(r) \to \left(1+\bar n_{\rm b}/\bar n_{\rm r}\right)^{-1} = \bar f_{\rm red}$ be reached. To get a rough estimate of when this will occur, let us ignore complications introduced by mass completeness, colour cuts, etc. and simply use a power law fit for the all-galaxy correlation function of our luminosity complete sample: $\xi(r)\sim (r/7{\rm Mpc})^{-1.85}$ \citep{zehavi+11}.  It is therefore not surprising that all our data sets, including the ones with $\rho=0.01$, show significant red fraction profiles at distances less than $\sim15$Mpc.
\item {\bf $f_{\rm red|rc}(r)=f_{\rm red|bc}(r)$ at large $r$ for all models:}\\ 
At large separations, it is reasonable to assume that halo bias is approximately linear and scale-independent. The 2-halo cross-correlations above can then be written as Fourier transforms of cross-power spectra that take the form $P^{\rm (2h)}_{\rm b(r),rc}(k) = P_{\rm lin}(k)\,\bar b_{\rm rc}\,\bar b_{\rm b(r)}(k)$ for $\xi^{\rm (2h)}_{\rm b(r),rc}(r)$, where $\bar b_{\rm b(r)}(k)$ is the number-weighted halo bias associated with all blue (red) galaxies while $\bar b_{\rm rc}$ is the corresponding term for red centrals alone, and $P_{\rm lin}(k)$ is the dark matter power spectrum in linear theory \citep[see, e.g.,][]{cs02}. (The scale dependence in the all-galaxy bias arises from the profiles of satellites within a halo, while the central-only term has no such scale dependence.) In the regime where both these terms are larger than unity, the scale independent $\bar b_{\rm rc}$ cancels in their ratio. A similar factor $\bar b_{\rm bc}$ cancels when constructing $f_{\rm red|bc}(r)$ as well, which means that we will have $f_{\rm red|rc}(r)\approx f_{\rm red|bc}(r)$ in this regime for any value of $\rho$. Indeed, we see this behaviour at large scales in Figure~\ref{fig:redfrac-trugrps}. 
\item {\bf All red fractions are larger than $\bar f_{\rm red}$:}\\ 
The red fractions (even in the absence of 2-halo conformity) approach the global mean from \emph{above} rather than below because the average bias factors of blue and red galaxies satisfy $\bar b_{\rm b} < \bar b_{\rm r}$. This may seem counter-intuitive since, on average, blue galaxies in a mass-complete sample are more luminous and therefore preferentially occupy more massive halos than red galaxies. However, the halo mass distribution of red galaxies is substantially broader than that of blue galaxies, with a longer tail at high masses that pushes the number-weighted red galaxy bias $\bar b_{\rm r}$ to be larger than $\bar b_{\rm b}$.
In the luminosity complete sample, this effect is further enhanced due to the inclusion of a large number of faint blue objects (predominantly centrals) which also live in low mass halos.
\item {\bf Increase of 2-halo conformity red fractions with $\rho$:}\\ 
The reason that the red fractions in the presence of 2-halo conformity increase with $\rho$ is the presence of halo assembly bias at large scales. In this case the bias factors above contain integrals of the form $\sim\int\der\ln m\int\der s\,p(s|m)\,b(m,s)\,\Phi_{\rm blue(red)}(m,s)$ which couple the concentration-dependence of halo bias to that of the red fraction in our model. The number weighting of bias means that small mass halos ($m_{\rm 200b}\sim10^{11.8}\Mh\sim0.25M_\ast$) will dominate the integral. If the bias $b(m,s)$ for such halos were monotonically increasing with $s$, then a straightforward calculation shows that a large positive value of $\rho$ would enhance the difference between $\bar b_{\rm b}$ and $\bar b_{\rm r}$ discussed previously, consequently increasing the red fractions. Although the actual behaviour of the bias for masses $\sim0.25M_\ast$ is non-monotonic and shows a minimum around $s=0$ \citep[see, e.g., Figure 4 of][]{wechsler+06}, the fact that the majority of galaxies in the mass-complete sample are red for any halo mass means that this trend does not change.
Interestingly, the amplitude of this signal appears stable against a possible weak mass dependence in the central red fraction (see Figure~\ref{fig:redfrac-conformity-2h}), suggesting that an enhanced value of $f_{\rm red|rc}(r) \simeq f_{\rm red|bc}(r)$ at $r\gtrsim8$Mpc as compared to the global average could be a 
robust indicator of 2-halo conformity.
\item {\bf $f_{\rm red|rc}(r) > f_{\rm red|bc}(r)$ for $r\lesssim5$Mpc for $\rho=0.99$ with 2-halo conformity:}\\ 
For $\rho=0.99$, the term ``red fraction'' is essentially the same as ``fraction with large halo concentration $s$'' (equation~\ref{newredfrac}). 
When computing red fractions we are then actually asking: what is the fraction of large-$s$ objects around (a) large-$s$ objects and (b) small-$s$ objects? If we now focus on separations $r\lesssim5$Mpc, we are directly studying a version of halo assembly bias in a range of scales (the so-called ``1-halo to 2-halo transition'' regime) where the previous assumptions regarding linear scale-independent biasing are no longer valid. 

Consider centrals in low mass halos first (e.g., the isolated singletons from earlier). Due to assembly bias, the ones with large $s$ will cluster strongly with other large-$s$ centrals of similar halo mass, while the small-$s$ centrals will be close to unbiased. The number weighting will ensure that this behaviour gives the dominant effect in the vicinity of small halo centrals, with little contribution from satellites and centrals in more massive halos. This will lead to a large difference in $f_{\rm red|rc}(r)$ and $f_{\rm red|bc}(r)$ at small distances, which must then smoothly match the large scale trend of $f_{\rm red|rc}(r)\approx f_{\rm red|bc}(r)$, consistent with what is seen in Figures~\ref{fig:redfrac-2hconf} and~\ref{fig:redfrac-cc}. 

Group centrals that reside in more massive halos are likely to have other high mass halos nearby due to larger average halo bias, so that assembly bias will still be active. However, contamination due to nearby singletons (which reside in less clustered, lower mass halos) and the fact that assembly bias reverses its trend around the global characteristic halo mass will mean that the strength of the difference in red fractions will be smaller here. This is consistent with the results in Figure~\ref{fig:redfrac-trugrps}.
\item {\bf $\rho=0.99$ without 2-halo conformity:}\\
For the model with $\rho=0.99$ and 2-halo conformity switched off, we expect that at large scales $f_{\rm red|rc}=f_{\rm red|bc}$ while at small scales $f_{\rm red|rc} \to 1$ and $f_{\rm red|bc}\to0$. The transition between these extremes will be dramatic but smooth due to averaging over halo mass, and should occur in the same ``1-halo to 2-halo transition'' regime discussed above. This is consistent with what we see in the right panel of Figure~\ref{fig:redfrac-2hconf}, where the transition occurs around $r\sim3$-$4$Mpc. In the left panel of Figure~\ref{fig:redfrac-trugrps} we essentially see only the large scale behaviour; presumably this is because we do not impose any isolation criterion in the group central measurement and therefore end up selecting very small groups (of which there are many). The isolation criterion in Figure~\ref{fig:redfrac-2hconf} on the other hand would select relatively large groups (in addition to the very high number of singletons which would however not contribute any 1-halo signal). 
This effect adds to \emph{and confuses} the transition regime 2-halo conformity effect discussed above, particularly when studying projected signals. We discuss this further in section~\ref{sec:discuss}.
\end{itemize}

\begin{figure}
\centering
\includegraphics[width=0.45\textwidth]{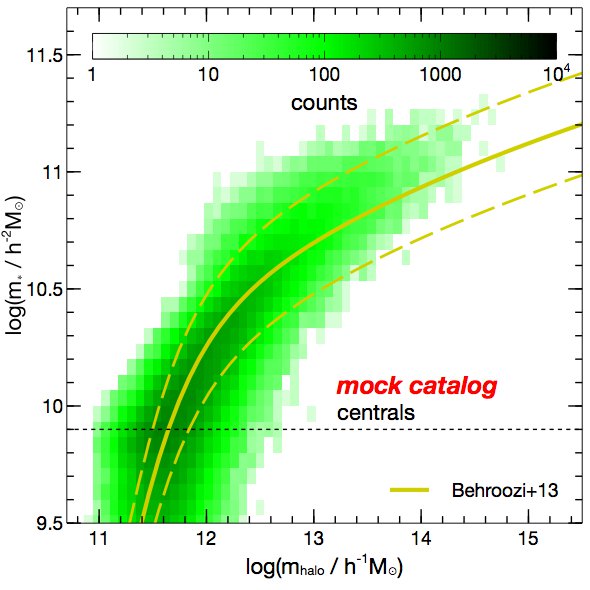}
\caption{Joint distribution of stellar mass of centrals and their parent halo mass. For comparison, the smooth yellow curves show the median stellar mass at fixed halo mass (solid curve) and its 1-sigma scatter (dashed curves) as calibrated by \citet{bwc13} at redshift $z=0$. The horizontal dotted line shows our mass completeness limit at $\log_{10}(m_\star)=9.9$.}
\label{fig:SHMR}
\end{figure}

Finally, the two panels of Figure~\ref{fig:redfrac-conformity-2h-smbins} are formatted identically to Figure~\ref{fig:redfrac-conformity-2h} and show that the large scale conformity signal in our mocks is essentially independent of the stellar mass of the isolated primary, an effect we attribute to the large scatter in the distribution of halo masses at fixed stellar mass as seen in Figure~\ref{fig:SHMR}.

\label{lastpage}


\end{document}